\documentclass[preprint]{JHEP3}
\usepackage{amsmath,amssymb}
\usepackage{stmaryrd}

\newcommand\fverb{\setbox\fverbbox=\hbox\bgroup\verb}
\newcommand\fverbdo{\egroup\medskip\noindent%
			\fbox{\unhbox\fverbbox}\ }
\newcommand\fverbit{\egroup\item[\fbox{\unhbox\fverbbox}]}
\newbox\fverbbox

\newcommand{\nlin}[2]{\href{http://xxx.lanl.gov/abs/nlin/#2}{\tt nlin.#1/#2}}

\title{On the reflection of magnon bound states}

\author{Niall MacKay$^a\,$\thanks{Email: nm15@york.ac.uk}
	~and Vidas Regelskis$^{a\,b\,}$\thanks{Email: vr509@york.ac.uk}\\
	$^a$ Department of Mathematics, University of York,\\
	$\;\;$ Heslington, York YO10 5DD, UK\\\\
	$^b$ Vilnius University Institute of Theoretical Physics and Astronomy,\\
	$\;\;$ Go\v{s}tauto 12, Vilnius 01108, Lithuania
}

\abstract{We investigate the reflection of two-particle
bound states of a free open string in the light-cone $AdS_{5}\times S^{5}$  string sigma model,
for large angular momentum $J=J_{56}$ and ending
on a D7 brane which wraps the entire $AdS_{5}$ and a maximal $S^3 \subset S^5$. We use
the superspace formalism to analyse fundamental
and two-particle bound states in the cases of supersymmetry-preserving and
broken-supersymmetry boundaries. We find the boundary $S$-matrices corresponding
to bound states both in the bulk and on the boundary.}

\begin{document}

\section{Introduction}

It has been recognized in recent years that the planar limit of $\mathcal{N}=4$
super Yang-Mills is integrable, and the $S$-matrix approach allows
us to successfully study the spectra of superstrings propagating freely
in $AdS_{5}\times S^{5}$ spacetime in the framework of the $AdS/CFT$
correspondence conjectured by Maldacena et al. \cite{Maldacena1}.
The $S$-matrix approach \cite{Arutyunov2,Arutyunov3,Staudacher1} was
first developed in the spin chain framework in the perturbative regime
of the gauge theory, where it allows one to conjecture the corresponding
(all-loop) Bethe equations describing
the asymptotic spectrum of the gauge theory \cite{Beisert1,Beisert3,Frolov1}.
Integrability allows one to find  exact expressions for the $S$-matrices by
requiring them to respect the underlying symmetries of the
model. It is well-known that the $S$-matrix for the fundamental excitations
in the bulk can be determined up to an overall (`dressing') phase
factor from just the centrally extended $\mathfrak{su}\left(2|2\right)$
symmetry \cite{Beisert1,Beisert2,Arutyunov4}, and the $S$-matrix so obtained
respects the Yang-Baxter equation (YBE) and a generalized physical unitarity condition.
The overall phase factor is severely constrained by  crossing
symmetry \cite{Janik1}. This non-analytic overall phase factor constitutes an
important feature of the string $S$-matrix and has been the subject of intensive
research \cite{Beisert5,Beisert6,Beisert7,Arutyunov5}.

In the limit of infinite light-cone momentum, in addition to the fundamental
particles, the spectrum of the string sigma model contains an infinite
tower of bound states \cite{Dorey2, Arutyunov7, Roiban1}. These manifest themselves as
poles of the multi-particle $S$-matrix built from the fundamental $S$-matrix
$S^{AA}$. More explicitly, $l$-particle bound states appear as
the tensor product of two 4$l$-dimensional atypical (short) totally symmetric multiplets
of the centrally extended $\mathfrak{su}\left(2|2\right)$ algebra
\cite{Dorey2,Dorey1,Beisert4}. This can be obtained from the $l$-fold
tensor product of the fundamental multiplets by projecting it onto the
totally symmetric component.

Construction of $S$-matrices for the bound states is more complicated,
as the $\mathfrak{su}\left(2|2\right)$ symmetry alone is no longer
 enough to determine the $S$-matrices uniquely. Further constraints
are required, either from the YBE or the underlying Yangian symmetry
\cite{Beisert4}; one can then determine general $l$-particle
bound state bulk $S$-matrices \cite{Arutyunov1,deLeeuw1,Arutyunov6}.
It is worth recalling that Yangians generically have some very nice
properties, particularly at the level of representation theory \cite{Bernard1,MacKay1}.
So the appearance of Yangian symmetry in the string context -- for example, via the universal R-matrix \cite{Ragoucy1,Torrielli1,Torrielli2} -- is
a very welcome feature.

As was shown in \cite{Arutyunov1}, the construction of the bound state
$S$-matrix relies on the observation that the $l$-particle bound state
representation $\mathcal{V}_{l}$ of the centrally extended $\mathfrak{su}\left(2|2\right)$
algebra may be realized on the space of homogeneous (super)symmetric
polynomials of degree $l$ depending on two bosonic and two fermionic
variables, $\omega_{a}$ and $\theta_{\alpha}$ respectively. Thus,
the representation space is identical to an irreducible short superfield
$\Phi^{l}\left(\omega,\theta\right)$. In this realization the algebra
generators are represented by differential operators $\mathbb{J}$
linear in variables $\omega_{a}$ and $\theta_{\alpha}$ with the
scattering coefficients being functions of the parameters describing
the representation. The introduction of a space $\mathcal{D}_{l}$
dual to $\mathcal{V}_{l}$, which may be realized as the space of
differential operators preserving the homogeneous gradation of $\Phi^{l}\left(\omega,\theta\right)$,
allows one to define the $S$-matrix as an element of
$$
\mbox{End}\left(\mathcal{V}^{l_{1}}\otimes\mathcal{V}^{l_{2}}\right)\approx\mathcal{V}^{l_{1}}
\otimes\mathcal{V}^{l_{2}}\otimes\mathcal{D}_{l_{1}}\otimes\mathcal{D}_{l_{2}}.
$$
Thus the $S$-matrix $S^{l_{1}l_{2}}$ may be written as a differential
operator of degree $l_{1}+l_{2}$ acting on the product of two superfields
$\Phi^{l_{1}}\left(\omega,\theta\right)$ and $\Phi^{l_{2}}\left(\omega,\theta\right)$.

The $S$-matrices $S^{AB}$ and $S^{BB}$ which describe the scattering
processes involving the fundamental multiplet $A$ and the two-particle
bound state multiplet $B$ were found in \cite{Arutyunov1}. The invariance conditions
for the latter only partially determine the scattering coefficients $a_{i}$, for if
the tensor product $\mathcal{V}^{l_{1}}\otimes\mathcal{V}^{l_{2}}$
has $m$ irreducible components, then $m-1$ coefficients $a_{i}$
together with an overall scale are left undetermined. The YBE
turns out to be sufficient to determine
the so-far-unrestricted $m-1$ coefficients $a_{i}$, leaving only the overall scale.

However, the underlying Yangian symmetry provides an alternative way of finding these
coefficients  \cite{Arutyunov6}, an approach which goes back to the inception of quantum groups \cite{Jimbo}.
This leads to a general strategy
for finding the $S$-matrices of higher order bound states \cite{Arutyunov6}, which is essential since the fusion
procedure does not work straightforwardly for AdS/CFT $S$-matrices \cite{Arutyunov1}. These
higher-order $S$-matrices play an important role in understanding the underlying
integrability and deriving the transfer matrices, Bethe ansatz equations,
and so on.

Very similar considerations apply to the analysis of the the spectra
of strings with open boundary conditions in the limit of infinite
light-cone momentum \cite{Young1,Dorey3,Galleas1,Palla1}. The reflection
of fundamental magnons from a boundary was considered in \cite{Young1,Palla1,Maldacena2},
while the reflection of magnon bound states was considered in \cite{Ahn2},
and the Yangian symmetry of an open string attached to the giant graviton
brane was recently exploited in \cite{Ahn1}. The first and key question is to determine
whether a particular boundary condition is integrable or not. It was shown
in \cite{Mann1,Mann2} that working to first-order in the \textquoteright{}t
Hooft coupling the D7 brane yields integrable boundary conditions
at least in the $\mathfrak{so}(6)$ sector. Further investigations
followed \cite{Swanson1,Takayama1,Takayama2}.

Deep in the bulk of an open spin chain
the theories are indistinguishable from pure $\mathcal{N}=4$,
so the symmetry arguments discussed above remain valid, and
the bulk $S$-matrix may be used without  modifications. The task is then to
 determine the reflection of magnons off
the end of the chain, where the residual symmetries of the boundary are
crucial in determining the structure of the reflection $K$-matrix. It
was shown in \cite{Young1} that the relative orientation between
the preferred $R$-charge of the vacuum and the spherical factor of the
brane worldvolume affect the symmetries preserved by the reflection,
and there are two inequivalent possibilities for reflection
from D5 and D7 branes; further, only in certain cases can the boundary itself
have an excitation attached to it. The nested coordinate Bethe
ansatz equations for D3 (maximal giant graviton) and D7 branes were recently proposed in \cite{Young2}.

In this paper we consider the so-called `$Z=0$ D7-brane'
system, in which the usual gauge/string correspondence in $AdS_{5}\times S^{5}$ has
a D7-brane wrapping the
entire $AdS_{5}$ and a maximal $S^{3}$ of the $S^{5}$ (defined by
setting $X^{5}=X^{6}=0$) with a $\mathcal{N}=2$ super Yang-Mills theory
living on it \cite{Young1}. The preferred R-charge is $J=J_{56}$
and we are considering states in which both $J$ and the classical
scaling dimension $\Delta$ are large, but keeping the difference
$\Delta-J$ finite. Hence the vacuum state $Z$ has $\Delta-J=0$
and the elementary magnons are the excitations with $\Delta-J=1$.

Our main goal is to fill a gap in the literature on the
reflection of bound states, particularly reflection from the D7
brane. The  $K$-matrices $K^{Aa}$ and $K^{A1}$ which describe
the scattering of the fundamental bulk multiplet $A$
off the fundamental boundary multiplet $a$ and singlet state
$1$ were found in \cite{Young1}, but the bound state
reflection matrices are unknown. Our goal is to find the $K$-matrices
$K^{Ba}$, $K^{Ab}$, $K^{Bb}$ and $K^{B1}$ using the the superspace
formalism presented in \cite{Arutyunov1}, where we denote the two
particle bound state multiplet on the boundary as $b$. We choose
the phase $\zeta$ increasing from left to right in accordance
with \cite{Beisert4,Young1,Maldacena2}, but in contrast to \cite{Arutyunov1}.
Thus we shall need to calculate the bulk bound state $S$-matrices independently
in order to check
that our $K$-matrices satisfy the YBE.

The outline of this paper is as follows. In section 2 we briefly recall
the  details of the scattering of fundamental magnons, in the
bulk and on the boundary. In section 3 we review the representation
algebra of magnon bound states, recall the superspace formalism
introduced in \cite{Arutyunov1}, and extend it to the boundary algebra.
In sections 4 and 5 we present the description of $S$- and $K$-
matrices in the superspace formalism. The results of our calculations,
which involve some large sets of coefficients,
are presented in appendices.

\section{Symmetries and fundamental representations}

In this section we shall briefly review the symmetries in the bulk
and on the boundary. We shall consider the superconformal algebra
$\mathfrak{psu}\left(4|4\right)$ of the $\mathcal{N}=4$ SYM in the bulk \cite{Arutyunov2} and
$\mathcal{N}=2$ SYM on the boundary \cite{Krucz1}. We build the scattering
theory whose vacuum state is the operator $\mathrm{tr}Z^{L}$ with $L\gg1$
for the closed boundary conditions and the operator
\begin{equation}
\epsilon_{j_{1},...,j_{N}}^{i_{1},...,i_{N}}Z_{i_{1}}^{j_{1}}...Z_{j_{N-1}}^{j_{L-1}}\left(\chi_{L}Z^{J}\chi_{R}\right)_{i_{N}}^{j_{N}},
\end{equation}
for the open boundary conditions, where $\chi_{L}$, $\chi_{R}$ are
the excitations living on the left and on the right boundaries and
$Z=\Phi_{5}+i\Phi_{6}$ is the vacuum reference state with the charge
under $\Delta-J$ being zero. All remaining states have $\Delta-J>0$.

\subsection{Bulk case}

The bulk superconformal algebra is
$\mathfrak{psu}\left(4|4\right)\cong\mathfrak{psu}\left(2|2\right)\times\widetilde{\mathfrak{psu}}\left(2|2\right)$.
We shall use the undotted and dotted indices to distinguish left and
right Lorentz generators $\mathbb{L}_{\alpha}^{\enskip\beta}\in\mathfrak{psu}\left(2|2\right)$,
$\tilde{\mathbb{L}}_{\dot{\alpha}}^{\enskip\dot{\beta}}\in\widetilde{\mathfrak{psu}}\left(2|2\right)$,
where
\begin{equation}
\alpha,\;\beta,...=+,-,\qquad\dot{\alpha},\;\dot{\beta},...=\dot{+},\dot{-},
\end{equation}
and $\mathbb{R}_{a}^{\enskip b}$, where
\begin{equation}
a,\; b,...=1,\;2,\;3,\;4,
\end{equation}
to denote R-symmetry generators. The same notation will be used for
supersymmetry generators $\mathbb{Q}_{\beta}^{\enskip b}$, $\mathbb{G}_{b}^{\enskip\beta}$
and $\tilde{\mathbb{Q}}_{\dot{\beta}}^{\enskip\dot{b}}$, $\tilde{\mathbb{G}}_{b}^{\enskip\dot{\beta}}$.
The supercharges transform canonically according the indices they carry:

\begin{eqnarray}
\left[\mathbb{L}_{\alpha}^{\enskip\beta},\mathbb{J}^{\gamma}\right]=\delta_{\alpha}^{\gamma}\mathbb{J}^{\beta}-\frac{1}{2}\delta_{\alpha}^{\beta}\mathbb{J}^{\gamma}, & \qquad & \left[\mathbb{L}_{\alpha}^{\enskip\beta},\mathbb{J}_{\gamma}\right]=-\delta_{\gamma}^{\beta}\mathbb{J}_{\alpha}+\frac{1}{2}\delta_{\alpha}^{\beta}\mathbb{J}_{\gamma},\nonumber \\
\left[\tilde{\mathbb{L}}_{\dot{\alpha}}^{\enskip\dot{\beta}},\mathbb{J}^{\dot{\gamma}}\right]=\delta_{\dot{\alpha}}^{\dot{\gamma}}\mathbb{J}^{\dot{\beta}}-\frac{1}{2}\delta_{\dot{\alpha}}^{\dot{\beta}}\mathbb{J}^{\dot{\gamma}}, & \qquad & \left[\tilde{\mathbb{L}}_{\dot{\alpha}}^{\enskip\dot{\beta}},\mathbb{J}_{\dot{\gamma}}\right]=-\delta_{\dot{\gamma}}^{\dot{\beta}}\mathbb{J}_{\dot{\alpha}}+\frac{1}{2}\delta_{\dot{\alpha}}^{\dot{\beta}}\mathbb{J}_{\dot{\gamma}},\nonumber \\
\left[\mathbb{R}_{a}^{\enskip b},\mathbb{J}^{c}\right]=\delta_{a}^{c}\mathbb{J}^{b}-\frac{1}{4}\delta_{a}^{b}\mathbb{J}^{c}, & \qquad & \left[\mathbb{R}_{a}^{\enskip b},\mathbb{J}_{c}\right]=-\delta_{c}^{b}\mathbb{J}_{a}+\frac{1}{4}\delta_{a}^{b}\mathbb{J}_{c}.
\end{eqnarray}
We shall be considering the subsectors $\mathfrak{psu}\left(2|2\right)$
and $\widetilde{\mathfrak{psu}}\left(2|2\right)$ of the whole symmetry
separately. The relevant algebra shall be centrally extended $\mathfrak{psu}\left(2|2\right)\ltimes\mathbb{R}^{3}$
which we shall denote as $\mathfrak{psu}\left(2|2\right)_{\mathcal{C}}$
\cite{Beisert1}. It is generated by the bosonic rotation generators $\mathbb{L}_{\alpha}^{\enskip\beta}$,
$\mathbb{R}_{a}^{\enskip b}$, the supersymmetry generators $\mathbb{Q}_{\beta}^{\enskip b}$,
$\mathbb{G}_{b}^{\enskip\beta}$, and three central charges $\mathbb{H}$,
$\mathbb{C}$ and $\mathbb{C}^{\dagger}$ obeying the relations
\begin{eqnarray}
\left\{ \mathbb{Q}_{\alpha}^{\enskip a},\mathbb{Q}_{\beta}^{\enskip b}\right\}  & = & \epsilon^{ab}\epsilon_{\alpha\beta}\mathbb{C},\nonumber \\
\left\{ \mathbb{G}_{a}^{\enskip\alpha},\mathbb{G}_{b}^{\enskip\beta}\right\}  & = & \epsilon^{\alpha\beta}\epsilon_{ab}\mathbb{C}^{\dagger},\nonumber \\
\left\{ \mathbb{Q}_{\alpha}^{\enskip a},\mathbb{G}_{b}^{\enskip\beta}\right\}  & = & \delta_{b}^{a}\mathbb{L}_{\beta}^{\enskip\alpha}+
\delta_{\beta}^{\alpha}\mathbb{R}_{b}^{\enskip a}+\delta_{b}^{a}\delta_{\beta}^{\alpha}\mathbb{H},
\end{eqnarray}
where $a,\; b,...=1,\;2$ and $\alpha,\;\beta,...=3,\;4$. We shall be using this notation throughout remaining of the paper.

The fundamental excitations propagating in the bulk transform in the
fundamental representation $\boxslash$ of the $\mathfrak{psu}\left(2|2\right)_{\mathcal{C}}$
and the bulk scattering matrix factors as a tensor product $S\otimes\tilde{S}$,
where each factor acts as
\begin{equation}
S/\tilde{S}:\:\boxslash\otimes\boxslash\rightarrow\boxslash\otimes\boxslash.
\end{equation}
The basis of the space consists a two of bosons $\phi_{a}$ transforming
as a doublet under $su\left(2\right)_{\mathbb{R}}$ and two fermions
$\psi_{\alpha}$ - a doublet under $su\left(2\right)_{\mathbb{L}}$.
The $\mathfrak{psu}\left(2|2\right)_{\mathcal{C}}$ supercharges act
on this basis in the following way:
\begin{eqnarray}
\mathbb{Q}_{\beta}^{\enskip b}\left|\phi_{a}\right\rangle =a\,\delta_{a}^{b}\left|\psi_{\beta}\right\rangle ,\qquad & \begin{alignedat}{1}\qquad\end{alignedat}
 & \mathbb{G}_{b}^{\enskip\beta}\left|\phi_{a}\right\rangle =c\,\epsilon^{\beta\,\alpha}\epsilon_{b\, a}\left|\psi_{\alpha}\right\rangle ,\nonumber \\
\mathbb{Q}_{\beta}^{\enskip b}\left|\psi_{\alpha}\right\rangle =b\,\epsilon^{b\, a}\epsilon_{\beta\,\alpha}\left|\phi_{a}\right\rangle , & \qquad &
\mathbb{G}_{b}^{\enskip\beta}\left|\psi_{\alpha}\right\rangle =d\,\delta_{\alpha}^{\beta}\left|\phi_{b}\right\rangle .
\end{eqnarray}
The coefficients $a,\: b,\: c,\: d$ respect the multiplet shortening
condition $ad-bc=1$ and are parametrized as \cite{Beisert2}
\begin{equation}
a=\sqrt{\frac{g}{2}}\eta,\quad b=\sqrt{\frac{g}{2}}\frac{i\zeta}{\eta}\left(\frac{x^{+}}{x^{-}}-1\right),\quad c=-\sqrt{\frac{g}{2}}\frac{\eta}{\zeta x^{+}},\quad d=-\sqrt{\frac{g}{2}}\frac{x^{+}}{i\eta}\left(\frac{x^{-}}{x^{+}}-1\right),
\end{equation}
where $\zeta$\footnote{Our definition of the parameter $\zeta$ should be replaced by $\zeta \mapsto i\zeta$ for consistency with \cite{Beisert1}.} is an overall phase factor, $\eta$ reflects the freedom
of the choice of spectral parameters $x^{\pm}$ obeying
\begin{equation}
{\rm e }^{ip}=\frac{x^{+}}{x^{-}},\qquad x^{+}+\frac{1}{x^{+}}-x^{-}-\frac{1}{x^{-}}=\frac{2i}{g}.
\end{equation}
We shall talk about the particular choice of $\zeta$ and $\eta$
in the next section.

The values of the central charges for the fundamental multiplet are
\begin{eqnarray}
C & = & ab = \frac{i}{2}g\left({\rm e}^{ip}-1\right){\rm e}^{2i\xi},\nonumber \\
C^{\dagger} & = & cd = -\frac{i}{2}g\left({\rm e}^{-ip}-1\right){\rm e}^{-2i\xi},\nonumber \\
H & = & ad+bc = \sqrt{1+4g^{2}\sin^{2}\frac{p}{2}},
\end{eqnarray}
where $H$ gives the energy-momentum dispersion relation of the states.

\subsection{Boundary case}

We shall consider the so-called `$Z=0$ D7-brane' system, where
the brane is wrapping the entire $AdS_{5}$ and a maximal $S^{3}\subset S^{5}$.
This case was nicely presented in \cite{Young1}. Here we shall briefly
review the properties of the configuration that are relevant to us.

The D7 brane is usually defined by setting $X_{5}=X_{6}=0$. This
choice breaks the $\mathfrak{so}\left(6\right)$ R-symmetry down to
$\mathfrak{so}\left(4\right)_{1234}\times\mathfrak{so}\left(2\right)_{56}$.
It was shown in \cite{Krucz1} that the presence of the D7-brane breaks
the half of the background supersymmetries that are left handed with
respect to the surviving $\mathfrak{so}\left(4\right)\subset\mathfrak{so}\left(6\right)$
and the surviving fields on the brane form the $\mathcal{N}=2$ hypermultiplet.
The choice of Bethe vacuum on the spin chain may further reduce the symmetries on the boundary. We shall consider
the standard $Z=X_{5}+iX_{6}$ bulk vacuum case. The preferred R-charge
$J\equiv J_{56}$ rotates the directions transverse to the brane and
preserves the full $\mathfrak{so}\left(4\right)_{1234}$ R-symmetry,
but breaks half of the supercharges, leaving the residual symmetry algebra on the
boundary to be $\mathfrak{su}\left(2\right)\times\mathfrak{su}\left(2\right)\times\widetilde{\mathfrak{psu}}\left(2|2\right)\ltimes\mathbb{R}^{3}$.
This means that fundamental matter fields transform in a $\left(1,\boxslash\right)$
representation of $\mathfrak{psu}\left(2|2\right)\times\widetilde{\mathfrak{psu}}\left(2|2\right)$.
It implies that the reflection matrix factors as a tensor product
\begin{equation}
K\otimes\tilde{K},
\end{equation}
where we have to consider two different reflection processes -- the
reflection from a supersymmetric boundary
\begin{equation}
\tilde{K}:\:\boxslash\otimes\boxslash\rightarrow\boxslash\otimes\boxslash,
\end{equation}
and reflection from a singlet state on the boundary
\begin{equation}
K:\:\boxslash\otimes1\rightarrow\boxslash\otimes1.
\end{equation}

The fundamental representation of the excitations on the boundary is parametrized
by the coefficients \cite{Maldacena2}
\begin{equation}
a_{B}=\sqrt{\frac{g}{2}}\eta_{B},\quad b_{B}=-\sqrt{\frac{g}{2}}\frac{i\zeta}{\eta_{B}},\quad c_{B}=-\sqrt{\frac{g}{2}}\frac{\eta_{B}}{\zeta x_{B}},\quad d_{B}=\sqrt{\frac{g}{2}}\frac{x_{B}}{i\eta_{B}},
\end{equation}
and the shortening (mass-shell) condition reads as
\begin{equation}
x_{B}+\frac{1}{x_{B}}=\frac{2i}{g}.
\end{equation}
The solution of the mass-shell condition
\begin{equation}
x_{B}=\frac{i}{g}\left(1+\sqrt{1+g^{2}}\right),
\end{equation}
is chosen to give a positive energy for the unexcited boundary state
\begin{equation}
\epsilon=ad+bc=\sqrt{1+g^{2}}.
\end{equation}
Note that the central charges $C$ and $C^{\dagger}$ are not conserved under the reflection,
otherwise  momentum would be preserved (only $p \mapsto p$ would be allowed) leaving no sensible notion of  reflection.
Rather the total values of all three central charges are preserved under  reflection
and the outgoing momentum is indeed $-p$ \cite{Young1}.

\section{Magnon bound states}

In this section we shall briefly discuss the representation structure of
magnon bound states and the superspace formalism introduced in \cite{Arutyunov1}.
In this framework the $S$- and $K$- matrices are naturally realized as
$\mathfrak{su}\left(2\right)\otimes\mathfrak{su}\left(2\right)$-invariant differential operators in the tensor product of two representations.

\subsection{The representation of bound states}

$l$-magnon bound states in the light-cone string theory on $AdS_{5}\times S^{5}$
are described by atypical totally symmetric representations of $\mathfrak{su}\left(2|2\right)_{\mathcal{C}}$.
The dimension of the representation is $2l|2l$ and it can be realized
on a graded vector space with the following basis:
\begin{itemize}
\item a tensor $\left|e_{a_{1}...a_{l}}\right\rangle $, symmetric in $a_{i}$
where $a_{i}=1,2$ are bosonic indices, giving $l+1$ bosonic states;
\item a tensor $\left|e_{a_{1}...a_{l-2}\alpha_{1}\alpha_{2}}\right\rangle $, symmetric in $a_{i}$ and skew-symmetric in $\alpha_{i}$
where $\alpha_{i}=3,\:4$ are fermionic indices, giving $l-1$ bosonic
states;
\item a tensor $\left|e_{a_{1}...a_{l-1}\alpha}\right\rangle $, symmetric in $a_{i}$,
giving $2l$ fermionic states.
\end{itemize}
The corresponding vector space is denoted as $\mathcal{V}^{l}\left(p,\zeta\right)$,
where $p$ and $\zeta$ in general are complex parameters of the representation.

The action of the bosonic generators of the symmetry algebra on the basis of the corresponding vector space is
\begin{eqnarray}
\mathbb{L}_{c}^{\enskip b}\left|e_{a_{1}...a_{l}}\right\rangle  & = & \delta_{a_{1}}^{b}\left|e_{c...a_{l}}\right\rangle +...+\delta_{a_{M}}^{b}\left|e_{a_{1}...c}\right\rangle -\frac{l}{2}\delta_{c}^{b}\left|e_{a_{1}...a_{l}}\right\rangle ,\nonumber \\
\mathbb{L}_{c}^{\enskip b}\left|e_{a_{1}...a_{l-2}\alpha_{1}\alpha_{2}}\right\rangle  & = & \delta_{a_{1}}^{b}\left|e_{c...a_{l-2}\alpha_{1}\alpha_{2}}\right\rangle +...+\delta_{a_{M}}^{b}\left|e_{a_{1}...c\alpha_{1}\alpha_{2}}\right\rangle -\frac{l-2}{2}\delta_{c}^{b}\left|e_{a_{1}...a_{l-2}\alpha_{1}\alpha_{2}}\right\rangle ,\nonumber \\
\mathbb{L}_{c}^{\enskip b}\left|e_{a_{1}...a_{l-1}\alpha}\right\rangle  & = & \delta_{a_{1}}^{b}\left|e_{c...a_{l-1}\alpha}\right\rangle +...+\delta_{a_{M}}^{b}\left|e_{a_{1}...c\alpha}\right\rangle -\frac{l-1}{2}\delta_{c}^{b}\left|e_{a_{1}...a_{l-1}\alpha}\right\rangle ;\\
\nonumber \\
\mathbb{R}_{\gamma}^{\enskip\beta}\left|e_{a_{1}...a_{l}}\right\rangle  & = & 0,\nonumber \\
\mathbb{R}_{\gamma}^{\enskip\beta}\left|e_{a_{1}...a_{l-2}\alpha_{1}\alpha_{2}}\right\rangle  & = & \delta_{\alpha_{1}}^{\beta}\left|e_{a_{1}...a_{l-2}\gamma\alpha_{2}}\right\rangle +\delta_{\alpha_{2}}^{\beta}\left|e_{a_{1}...c\alpha_{1}\alpha_{2}}\right\rangle -\delta_{\gamma}^{\beta}\left|e_{a_{1}...a_{l-2}\alpha_{1}\alpha_{2}}\right\rangle ,\nonumber \\
\mathbb{R}_{\gamma}^{\enskip\beta}\left|e_{a_{1}...a_{l-1}\alpha}\right\rangle  & = & \delta_{\alpha}^{b}\left|e_{a_{1}...a_{l-1}\gamma}\right\rangle -\frac{1}{2}\delta_{\gamma}^{\beta}\left|e_{a_{1}...a_{l-1}\alpha}\right\rangle ;
\end{eqnarray}
while the action of the supersymmetric generators has the form
\begin{eqnarray}
\mathbb{Q}_{\beta}^{\enskip b}\left|e_{a_{1}...a_{l}}\right\rangle  & = & a_{1}^{l}\left(\delta_{a_{1}}^{b}\left|e_{a_{2}...a_{l}\alpha}\right\rangle +...+\delta_{a_{l}}^{b}\left|e_{a_{1}...a_{l-1}\alpha}\right\rangle \right),\nonumber \\
\mathbb{Q}_{\beta}^{\enskip b}\left|e_{a_{1}...a_{l-2}\alpha_{1}\alpha_{2}}\right\rangle  & = & b_{2}^{l}\epsilon^{b\, a_{l-1}}\left(\epsilon_{\beta\,\alpha_{1}}\left|e_{c...a_{l-1}\alpha_{2}}\right\rangle -\epsilon_{\beta\,\alpha_{2}}\left|e_{c...a_{l-1}\alpha_{1}}\right\rangle \right),\nonumber \\
\mathbb{Q}_{\beta}^{\enskip b}\left|e_{a_{1}...a_{l-1}\alpha}\right\rangle  & = & b_{1}^{l}\epsilon^{b\, a_{l}}\epsilon_{\beta\,\alpha}\left|e_{c...a_{l}}\right\rangle +a_{2}^{l}\left(\delta_{a_{1}}^{b}\left|e_{a_{2}...a_{l-1}\beta\alpha}\right\rangle +...+\delta_{a_{l-1}}^{b}\left|e_{a_{1}...a_{l-2}\beta\alpha}\right\rangle \right);\nonumber \\
\\
\mathbb{G}_{b}^{\enskip\beta}\left|e_{a_{1}...a_{l}}\right\rangle  & = & c_{1}^{l}\epsilon^{\beta\,\alpha}\left(\epsilon_{b\, a_{1}}\left|e_{a_{2}...a_{l}\alpha}\right\rangle +...+\epsilon_{b\, a_{l}}\left|e_{a_{1}...a_{l-1}\alpha}\right\rangle \right),\nonumber \\
\mathbb{G}_{b}^{\enskip\beta}\left|e_{a_{1}...a_{l-2}\alpha_{1}\alpha_{2}}\right\rangle  & = & d_{2}^{l}\left(\delta_{\alpha_{1}}^{\beta}\left|e_{a_{1}...a_{l-2}b\alpha_{2}}\right\rangle -\delta_{\alpha_{2}}^{\beta}\left|e_{a_{1}...a_{l-2}b\alpha_{1}}\right\rangle \right),\nonumber \\
\mathbb{G}_{b}^{\enskip\beta}\left|e_{a_{1}...a_{l-1}\alpha}\right\rangle  & = & d_{1}^{l}\delta_{\alpha}^{\beta}\left|e_{a_{1}...a_{l-1}b}\right\rangle +c_{2}^{l}\epsilon^{\beta\,\gamma}\left(\epsilon_{b\, a_{1}}\left|e_{a_{2}...a_{l-1}\gamma\alpha}\right\rangle +...+\epsilon_{b\, a_{l-1}}\left|e_{a_{1}...a_{l-2}\gamma\alpha}\right\rangle \right).\nonumber \\
\end{eqnarray}

The parameters $a_{i}^{l}$, $b_{i}^{l}$, $c_{i}^{l}$, $d_{i}^{l}$
are representation-dependent and may be determined by requiring that they
 respect the centrally extended $\mathfrak{su}\left(2|2\right)_{c}$
algebra \cite{Beisert1}, which imposes the  constraints
\begin{eqnarray}
b_{1}^{l}d_{2}^{l}=b_{2}^{l}d_{1}^{l}, & \qquad & c_{1}^{l}d_{2}^{l}=c_{2}^{l}d_{1}^{l},\nonumber \\
a_{1}^{l}d_{1}^{l}-b_{1}^{l}c_{1}^{l}=1, & \qquad & a_{2}^{l}d_{2}^{l}-b_{2}^{l}c_{2}^{l}=1.
\end{eqnarray}
The central charges obey the shortening condition
\begin{equation}
\mathbb{H}_{l}^{2}-4\mathbb{C}_{l}\mathbb{C}_{l}^{\dagger}=1.
\end{equation}
The eigenvalue of $\mathbb{H}_{l}$ depends explicitly on the bound state number $l$ in the following way
\begin{equation}
H_{l} = \sqrt{l^{2}+4g^{2}\sin^{2}\frac{p}{2}}=l\sqrt{1+4\left(\frac{g}{l}\right)^{2}\sin^{2}\frac{p}{2}},
\end{equation}
where $\frac{g}{l}$ may be called the effective coupling constant for an $l$-magnon bound state.
In this way the values of $\mathbb{C}_{l}$ and $\mathbb{C}_{l}^{\dagger}$ may be defined to depend explicitly on $l$ by setting
\begin{equation}
C_{l} = l\frac{i}{2}\frac{g}{l}\left({\rm e}^{ip}-1\right){\rm e}^{2i\xi},\qquad
C_{l}^{\dagger} = -l\frac{i}{2}\frac{g}{l}\left({\rm e}^{-ip}-1\right){\rm e}^{-2i\xi}.
\end{equation}
This yields the usual definition of central charges in terms of representation parameters,
\begin{align}
 \frac{C_{l}}{l}=a_{1}^{l}d_{1}^{l}=a_{2}^{l}d_{2}^{l},\qquad \frac{C_{l}^{\dagger}}{l}=c_{1}^{l}d_{1}^{l}=c_{2}^{l}d_{2}^{l},\nonumber \\
 \frac{H_{l}}{l}=\left(a_{1}^{l}d_{1}^{l}+b_{1}^{l}c_{1}^{l}\right)=\left(a_{2}^{l}d_{2}^{l}+b_{2}^{l}c_{2}^{l}\right),\;
\end{align}
implying that it is always possible to choose  $a_{i}^{l}$, $b_{i}^{l}$,
$c_{i}^{l}$, $d_{i}^{l}$ so that
\begin{equation}
a_{1}^{l}=a_{2}^{l}\equiv a,\quad b_{1}^{l}=b_{2}^{l}\equiv b,\quad c_{1}^{l}=c_{2}^{l}\equiv c,\quad d_{1}^{l}=d_{2}^{l}\equiv d
\end{equation}
and thereby obtaining the the convenient parametrization
\begin{equation}
a=\sqrt{\frac{g}{2l}}\eta,\quad b=\sqrt{\frac{g}{2l}}\frac{i\zeta}{\eta}\left(\frac{x^{+}}{x^{-}}-1\right),\quad c=-\sqrt{\frac{g}{2l}}\frac{\eta}{\zeta x^{+}},\quad d=-\sqrt{\frac{g}{2l}}\frac{x^{+}}{i\eta}\left(\frac{x^{-}}{x^{+}}-1\right),
\end{equation}
where $\zeta={\rm e}^{2i\xi}$ and the spectral parameters $x^{\pm}$ respect
the mass-shell condition of the $l$-magnon bound state,
\begin{equation}
x^{+}+\frac{1}{x^{+}}-x^{-}-\frac{1}{x^{-}}=i\frac{2l}{g}.\label{shortening}
\end{equation}
The conservation of central charges on the sum of an $l$- and an $m$-magnon bound state requires
\begin{eqnarray}
C_{l+m} & = & C_{l}+C_{m}\nonumber \\
 & = & \frac{i}{2}g\left({\rm e}^{ip_{1}}-1\right){\rm e}^{2i\xi_{1}}+\frac{i}{2}g\left({\rm e}^{ip_{2}}-1\right){\rm e}^{2i\xi_{2}}\nonumber \\
 & = & \frac{i}{2}g\left({\rm e}^{ip}-1\right){\rm e}^{2i\xi_{0}},\label{C}
\end{eqnarray}
which is satisfied by setting the total momentum $p=p_{1}+p_{2}$ and $\xi_{1}\equiv\xi_{0}$,
$\xi_{2}\equiv\xi_{0}+\frac{p_{1}}{2}$. The same holds for $\mathbb{C}_{l}^{\dagger}$.

The unitarity condition implies $d^{*}=a$, $c^{*}=b$, hence
\begin{eqnarray}
\eta  =   \left[\frac{1}{i\eta}\left(x^{+}-x^{-}\right)\right]^{*}&=&-\frac{i}{\eta^{*}}{\rm e}^{i\varphi}\left(x^{+}-x^{-}\right),\nonumber \\[0.07in]
\frac{i\zeta}{\eta}\left(\frac{x^{+}}{x^{-}}-1\right)  =  -\left[\frac{\eta}{\zeta x^{+}}\right]^{*}&=&-\frac{\eta^{*}}{\zeta^{*}{\rm e}^{i\varphi}x^{-}}.
\end{eqnarray}
Here we have used the relation $\left(x^{\pm}\right)^{*}={\rm e}^{i\varphi}x^{\mp}$,
where the phase factor ${\rm e}^{i\varphi}$ represents the freedom to choose
the basis for $x^{\pm}$. These relations give
\begin{eqnarray}
\left|\eta\right|^{2} & = & i{\rm e}^{i\varphi}\left(x^{-}-x^{+}\right)\zeta,\nonumber \\[0.07in]
\eta & = & {\rm e}^{i\xi}{\rm e}^{i\frac{\varphi}{2}}\sqrt{i\left(x^{-}-x^{+}\right)}.\label{eta}
\end{eqnarray}
Unitarity also implies that parameters $\xi$ and $\varphi$ must be real. Constraints on $\xi$ were derived above,
while $\varphi$ may be chosen to acquire any value. The value $\varphi = 0$ is commonly used for the fundamental representation \cite{Maldacena2,Beisert2,Young1},
while the value $\varphi = \frac{p}{2}$ is prefered for the case of bound states \cite{Arutyunov7}.

The same considerations may be trivially extended for the representation of the boundary multi-magnon bound states.
Thus the boundary representation of $l$-magnon bound states is described by the parameters
\begin{equation}
a_{B}=\sqrt{\frac{g}{2l}}\eta_{B},\quad b_{B}=-\sqrt{\frac{g}{2l}}\frac{i\zeta}{\eta_{B}},\quad c_{B}=
-\sqrt{\frac{g}{2l}}\frac{\eta_{B}}{\zeta x_{B}},\quad d_{B}=\sqrt{\frac{g}{2l}}\frac{x_{B}}{i\eta_{B}},
\end{equation}
and the shortening condition reads as
\begin{equation}
x_{B}+\frac{1}{x_{B}}=i\frac{2l}{g}.\label{shortening_b}
\end{equation}
The unitarity condition for the boundary representation gives
\begin{equation}
\left|\eta_{B}\right|^{2} = -ix_{B},
\end{equation}
implying that boundary spectral parameter $x_{B}$ is purely imaginary.

\subsection{Superspace representation of $\mathfrak{su}\left(2|2\right)_{\mathcal{C}}$}

For a convenient description of the $S$-matrix a $2l|2l$ graded vector
space of monomials of degree $l$ of two bosonic $\omega_{a}$, $a=1,\:2$,
and two fermionic variables $\theta_{\alpha}$, $\alpha=3,\:4$ may
be introduced \cite{Arutyunov1}. Then any homogeneous symmetric polynomial
of degree $l$ can be expressed as
$$
\Phi_{l}\left(\omega,\theta\right)=\phi^{a_{1}..a_{l}}\omega_{a_{1}}\cdot\cdot\cdot\omega_{a_{l}}+\phi^{a_{1}...a_{l-1}\alpha}\omega_{a_{1}}
\cdot\cdot\cdot\omega_{a_{l-1}}\theta_{\alpha}+\phi^{a_{1}...a_{l-2}\alpha_{1}\alpha_{2}}\omega_{a_{1}}\cdot\cdot\cdot\omega_{a_{l-2}}\theta_{\alpha_{1}}\theta_{\alpha_{2}}.
$$
The basis of monomials is related to the vector space of magnon bound states
by
\begin{eqnarray}
\left|m,n,\mu,\nu\right\rangle  & = & N_{mn\mu\nu}\,\omega_{1}^{m}\omega_{2}^{n}\theta_{3}^{\mu}\theta_{4}^{\nu},\label{vect}
\end{eqnarray}
where $m,\: n\geq0$, $\mu,\:\nu=0,\:1$, $m+n+\mu+\nu=l$ and the
normalization constant is \cite{Arutyunov1}
\begin{eqnarray}
N_{mn\mu\nu} & = & \left(\frac{\left(l-1\right)!}{m!\: n!}\right)^{1/2}.
\end{eqnarray}
This basis is assumed to be orthogonal
\begin{equation}
\left\langle a,b,\alpha,\beta\big|m,n,\mu,\nu\right\rangle =\delta_{am}\delta_{bn}\delta_{\alpha\mu}\delta_{\beta\nu}.\label{orth}
\end{equation}
The hermitian conjugate operators are expressed as
\begin{equation}
\left(\omega_{a}\right)^{\dagger} = \frac{\partial}{\partial\omega_{a}},\qquad\left(\theta_{\alpha}\right)^{\dagger} = \frac{\partial}{\partial\theta_{\alpha}}
\end{equation}
and are considered to be real.

In this representation the centrally extended $\mathfrak{su}\left(2|2\right)$
generators are realized as the differential operators
\begin{eqnarray}
\mathbb{L}_{a}^{\enskip b} & = & \omega_{a}\frac{\partial}{\partial\omega_{b}}-\frac{1}{2}\delta_{a}^{b}\omega_{c}\frac{\partial}{\partial\omega_{c}},\nonumber \\
\mathbb{R}_{\alpha}^{\enskip\beta} & = & \theta_{\alpha}\frac{\partial}{\partial\theta_{\beta}}-\frac{1}{2}\delta_{\alpha}^{\beta}\theta_{\gamma}\frac{\partial}{\partial\theta_{\gamma}},\nonumber \\
\mathbb{Q}_{\alpha}^{\enskip a} & = & a\theta_{\alpha}\frac{\partial}{\partial\omega_{a}}+b\epsilon^{ab}\epsilon_{\alpha\beta}\omega_{b}\frac{\partial}{\partial\theta_{\beta}},\nonumber \\
\mathbb{G}_{a}^{\enskip\alpha} & = & c\epsilon_{ab}\epsilon^{\alpha\beta}\theta_{\beta}\frac{\partial}{\partial\omega_{b}}+d\omega_{a}\frac{\partial}{\partial\theta_{\alpha}},
\end{eqnarray}
while the central charges are
\begin{eqnarray}
\mathbb{C} & = & ab\left(\omega_{a}\frac{\partial}{\partial\omega_{a}}+\theta_{\alpha}\frac{\partial}{\partial\theta_{\alpha}}\right),\nonumber \\
\mathbb{C}^{\dagger} & = & cd\left(\omega_{a}\frac{\partial}{\partial\omega_{a}}+\theta_{\alpha}\frac{\partial}{\partial\theta_{\alpha}}\right),\nonumber \\
\mathbb{H} & = & \left(ad+bc\right)\left(\omega_{a}\frac{\partial}{\partial\omega_{a}}+\theta_{\alpha}\frac{\partial}{\partial\theta_{\alpha}}\right).
\end{eqnarray}

\subsection{$S$- and $K$- matrices in superspace formalism}

The $S$-matrix in superspace is realized as a differential operator
acting on the tensor product of two vector spaces
\begin{equation*}
\mathcal{V}^{M}\left(p_{1},\zeta_{1}\right)\otimes\mathcal{V}^{N}\left(p_{2},\zeta_{2}\right)\sim\mathcal{V}^{M}\left(p_{1},1\right)\otimes\mathcal{V}^{N}\left(p_{2},{\rm e}^{ip_{1}}\right)\sim\mathcal{V}^{M}\left(p_{1},{\rm e}^{ip_{2}}\right)\otimes\mathcal{V}^{N}\left(p_{2},1\right).
\end{equation*}
The unitarity condition implies that there are two possible equivalent
choices of phase factors, $\zeta_{1}=\zeta,\;\zeta_{2}=\zeta {\rm e}^{ip_{1}}$
and $\zeta_{1}=\zeta {\rm e}^{ip_{2}},\;\zeta_{2}=\zeta$. We define
the $S$-matrix as an intertwining operator
\begin{equation*}
S\left(p_{1},p_{2}\right):\quad\mathcal{V}^{M}\left(p_{1},\zeta\right)\otimes\mathcal{V}^{N}\left(p_{2},\zeta {\rm e}^{ip_{1}}\right)\rightarrow\mathcal{V}^{M}\left(p_{1},\zeta {\rm e}^{ip_{2}}\right)\otimes\mathcal{V}^{N}\left(p_{2},\zeta\right),
\end{equation*}
where the phase $\zeta$ increases from left to right.
Our definition is consistent with that of Beisert et al.%
\footnote{The usual $S$-matrix in the physical space $S^{phys}$ is related to the $S$-matrix
in the superspace $S^{super}$ as $S^{phys} = \mathcal{P} \cdot S^{super}$, where $\mathcal{P}$ is an ordinary graded permutation operator.}
but differs from that of Arutyunov et al., who choose the phase $\zeta$ to increase from right to left
\begin{equation*}
S\left(p_{1},p_{2}\right):\quad\mathcal{V}^{M}\left(p_{1},\zeta {\rm e}^{ip_{2}}\right)\otimes\mathcal{V}^{N}\left(p_{2},\zeta\right)\rightarrow\mathcal{V}^{M}\left(p_{1},\zeta\right)\otimes\mathcal{V}^{N}\left(p_{2},\zeta {\rm e}^{ip_{1}}\right).
\end{equation*}

In this superspace formalism the $S$-matrix may be viewed as an element
of\begin{equation*}
End\left(\mathcal{V}^{M}\otimes\mathcal{V}^{N}\right)\approx\mathcal{V}^{M}\otimes\mathcal{V}^{N}\otimes\mathcal{D}_{M}\otimes\mathcal{D}_{N},
\end{equation*}
where $\mathcal{D}_{M}$ is the vector space dual to $\mathcal{V}^{M}$.
The dual vector space is realized as the space of polynomials of degree
$M$ of the differential operators $\frac{\partial}{\partial\omega_{a}}$
and $\frac{\partial}{\partial\theta_{\alpha}}$ with a natural pairing
between $\mathcal{D}_{M}$ and $\mathcal{V}^{M}$ induced by the
relations $\frac{\partial}{\partial\omega_{a}}\omega_{b}=\delta_{b}^{a}$
and $\frac{\partial}{\partial\theta_{\alpha}}\theta_{\beta}=\delta_{\beta}^{\alpha}$.
Thus the $S$-matrix may be represented as
\begin{equation}
S\left(p_{1},p_{2}\right)=\sum_{i} a_{i}\left(p_{1},p_{2}\right)\Lambda_{i},
\end{equation}
where $\Lambda_{i}$ span a complete basis of differential operators
invariant under the $\mathfrak{su}\left(2\right)\oplus\mathfrak{su}\left(2\right)$
algebra and $a_{i}\left(p_{1},p_{2}\right)$ are $S$-matrix components.
The exact expression for $\Lambda_{i}$ for various $S$-matrices are
given in \cite{Arutyunov1}.

Following the above analysis, we define the $K$-matrix describing the
reflection of the bulk magnons from the boundary states as an operator
acting on the tensor space in the following way:
\begin{equation*}
K\left(p,q\right):\quad\mathcal{V}^{M}\left(p,\zeta\right)\otimes\mathcal{V}^{N}\left(q,\zeta {\rm e}^{ip}\right)
\rightarrow\mathcal{V}^{M}\left(-p,\zeta\right)\otimes\mathcal{V}^{N}\left(q,\zeta {\rm e}^{2ip}\right),
\end{equation*}
where $p$ is the momentum of the bulk state and $q$ is some parameter describing the
 boundary state. Hence, the reflection matrix can be represented
as
\begin{equation}
K\left(p,q\right) = \sum_{i} k_{i}\left(p,q\right)\Lambda_{i},
\end{equation}
where $\Lambda_{i}$ have the same form as for bulk $S$-matrices.

\section{$S$-matrices}

\subsection*{$S$-matrix $S^{AA}$}

We define the $S$-matrix $S^{AA}$ as an intertwining operator
\begin{equation*}
S^{AA}:\quad\mathcal{V}^{1}\left(p_{1},\zeta\right)\otimes\mathcal{V}^{1}\left(p_{2},\zeta {\rm e}^{ip_{1}}\right)
\rightarrow\mathcal{V}^{1}\left(p_{1},\zeta {\rm e}^{ip_{2}}\right)\otimes\mathcal{V}^{1}\left(p_{2},\zeta\right),
\end{equation*}
where $\mathcal{V}^{1}\otimes\mathcal{V}^{1}=\mathcal{W}^{2}$ is
isomorphic to a typical (long) multiplet of dimension 16. Thus the $S$-matrix
is described as the second-order differential operator
\begin{equation}
S^{AA} = \sum_{i=1}^{10}a_{i}\,\Lambda_{i},
\end{equation}
where the differential operators $\Lambda_{i}$ are given in (4.5)
of section 4.5 of \cite{Arutyunov1}. The $S$-matrix $S^{AA}$ coefficients
$a_{i}$ may be determined uniquely up to a overall constant using the
symmetry algebra, and the full expression in our basis is given
in the appendix. It was also shown that $S^{AA}$ respects Yangian
symmetry \cite{Beisert4}.

\subsection*{$S$-matrix $S^{AB}$}

We define the $S$-matrix $S^{AB}$ as an intertwining operator
\begin{equation*}
S^{AB}:\quad\mathcal{V}^{1}\left(p_{1},\zeta\right)\otimes\mathcal{V}^{2}\left(p_{2},\zeta {\rm e}^{ip_{1}}\right)
\rightarrow\mathcal{V}^{1}\left(p_{1},\zeta {\rm e}^{ip_{2}}\right)\otimes\mathcal{V}^{2}\left(p_{2},\zeta\right),
\end{equation*}
where $\mathcal{V}^{1}\otimes\mathcal{V}^{2}=\mathcal{W}^{3}$ is
isomorphic to a long multiplet of dimension 32. Thus the $S$-matrix
is described as the third-order differential operator
\begin{equation}
S^{AB} = \sum_{i=1}^{19}a_{i}\,\Lambda_{i},
\end{equation}
where the $\Lambda_{i}$ are given in section
6.1.1 of \cite{Arutyunov1}. The reflection coefficients $a_{i}$ may be determined uniquely up to a overall
constant using the symmetry algebra \cite{Arutyunov1}\footnote{We found two typos in \cite{Arutyunov1} in the coefficients of $S^{AB}$
listed in 6.1.2. There should be a numerator $\left(x_{1}^{+}-y_{2}^{+}\right)$
instead of $\left(x_{1}^{-}-y_{2}^{+}\right)$ in $a_{13}$ and a
numerator $\left(1-y_{2}^{-}x_{1}^{+}\right)$ instead of $\left(1-y_{2}^{-}x_{1}^{-}\right)$
in $a_{14}$. These typos were noted also in \cite{Ahn1}.};
the exact expression in our basis
is again given in the appendix.
It was also shown that $S^{AB}$ respects Yangian symmetry \cite{deLeeuw1,Arutyunov6}.

\subsection*{$S$-matrix $S^{BB}$}

We define the $S$-matrix $S^{BB}$ as an intertwining operator
\begin{equation*}
S^{BB}:\quad\mathcal{V}^{2}\left(p_{1},\zeta\right)\otimes\mathcal{V}^{2}\left(p_{2},\zeta {\rm e}^{ip_{1}}\right)
\rightarrow\mathcal{V}^{2}\left(p_{1},\zeta {\rm e}^{ip_{2}}\right)\otimes\mathcal{V}^{2}\left(p_{2},\zeta\right),
\end{equation*}
where $\mathcal{V}^{2}\otimes\mathcal{V}^{2}=\mathcal{W}^{2}\oplus\mathcal{W}^{4}$
are long multiplets of dimension 16 and 48 respectively. Thus the $S$-matrix is described as the following fourth-order differential operator
\begin{equation}
S^{BB} = \sum_{i=1}^{48}a_{i}\,\Lambda_{i},
\end{equation}
with $\Lambda_{i}$ as in section
6.2.1 of \cite{Arutyunov1}. It was shown in \cite{Arutyunov1} that the Lie
superalgebra alone is not enough to fix all the coefficients $a_{i}$, and the YBE is required.
This is the consequence of the decomposition of tensor product
$\mathcal{V}^{2}\otimes\mathcal{V}^{2}$ being a sum of two long multiplets
$\mathcal{W}^{2}\oplus\mathcal{W}^{4}$. Therefore the $S$-matrix $S^{BB}$ may be formally divided into two parts
\begin{equation}
S^{BB} = S^{BB}_{1}+q S^{BB}_{2},
\end{equation}
where $q$ is a single parameter which is not determined by the symmetry algebra \cite{Arutyunov1}.
In the case of higher-order multi-magnon bound state $S$-matrices $S^{MN}$, there are precisely $m-1$ parameters that cannot be
determined by the symmetry algebra alone, where $m$ is the number of long multiplets
in the decomposition of tensor product $\mathcal{V}^{M}\otimes\mathcal{V}^{N}$%
\footnote{The multiplet decomposition formula is given in p.19 of \cite{Arutyunov1}.
Long and short multiplets of $\mathfrak{su}\left(2|2\right)$ have been studied in \cite{Beisert1}.}.
Hence, in addition to the Lie superalgebra, the YBE or Yangian symmetry is required
to find $S^{MN}$ for $M, N \geq 2$ \cite{deLeeuw1,Arutyunov6}.

Precise expressions  for the scattering coefficients $a_{1},...,a_{48}$ of the $S$-matrix $S^{BB}$
were found in \cite{Arutyunov1}%
\footnote{We found a typo in \cite{Arutyunov1} in the coefficient $a_{41}$ of
$S^{BB}$ listed in 6.2.2. There should be a numerator $\tilde{\eta}_{2}$
instead of $\tilde{\eta}_{2}^{2}$. It is easy to see this by comparing
the expressions of $a_{41}$ with $a_{42}$ and $\Lambda_{41}$ with
$\Lambda_{42}$ from the section 6.2.1.%
}, by using the superalgebra together with YBE.
It is interesting to note that $a_{45},...,a_{48}$ were found to be zero:
the scattering channels $\Lambda_{44},...,\Lambda_{48}$ are forbidden,
and $a_{45}=...=a_{48}=0$ independently of the choice of parametrization
of  $a$, $b$, $c$ and $d$. In fact $a_{45}=...=a_{48}=0$
may be used as additional constraints and one can then obtain all the other scattering coefficients $a_{1},...,a_{44}$
using the symmetry algebra alone, a fact which we expect will be explained by a deeper understanding of the underlying Yangian symmetry.

The full expressions for the coefficients $a_{1},...,a_{48}$ of the $S$-matrix $S^{BB}$ in our basis are again given in the appendix.

\section{$K$-matrices}

\subsection*{$K$-matrix $K^{Aa}$}

We define the $K$-matrix $K^{Aa}$ as an intertwining operator
\begin{equation*}
K^{Aa}:\quad\mathcal{V}^{1}\left(p,\zeta\right)\otimes\mathcal{V}^{1}\left(q,\zeta {\rm e}^{ip}\right)
\rightarrow\mathcal{V}^{1}\left(-p,\zeta\right)\otimes\mathcal{V}^{1}\left(q,\zeta {\rm e}^{-ip}\right),
\end{equation*}
corresponding to the tilded factor of the complete reflection $K$-matrix
\begin{equation*}
\tilde{K}:\:\boxslash\otimes\boxslash\rightarrow\boxslash\otimes\boxslash.
\end{equation*}
The $K$-matrix $K^{Aa}$ is given by the second-order differential operator
\begin{equation}
K^{Aa} = \sum_{i=1}^{10}k_{i}\,\Lambda_{i},
\end{equation}
with the $\Lambda_{i}$  as in
(4.5) of section 4.1 of \cite{Arutyunov1}. The symmetry algebra fixes
the values of the coefficients $k_{i}$ uniquely up to a constant; again we reserve the full
expression to the appendix. These coefficients were derived in \cite{Young1}.

\subsection*{$K$-matrix $K^{A1}$}

We define the $K$-matrix $K^{A1}$ as an intertwining operator
\begin{equation*}
K^{Aa}:\quad\mathcal{V}^{1}\left(p,\zeta\right)\otimes1\rightarrow\mathcal{V}^{1}\left(-p,\zeta\right)\otimes1,
\end{equation*}
corresponding to the untilded factor of the complete reflection $K$-matrix
\begin{equation*}
\tilde{K}:\:\boxslash\otimes1\rightarrow\boxslash\otimes1,
\end{equation*}
The $K$-matrix $K^{A1}$ may be expressed as a sum of diagonal first-order differential operators
\begin{equation}
K^{A1} = \sum_{i=1}^{2}k_{i}\,\Lambda_{i},
\end{equation}
where
\begin{equation}
\Lambda_{1}=\omega_{a}^{1}\frac{\partial}{\partial\omega_{a}^{1}},\qquad\Lambda_{2}=\theta_{\alpha}^{1}\frac{\partial}{\partial\theta_{\alpha}^{1}}.
\end{equation}
As was shown in \cite{Young1}, the symmetry algebra alone is not
enough to fix $k_{1}$ and $k_{2}$. Using the boundary Yang-Baxter equation (bYBE)
\begin{equation*}
S^{AA}(p_{1},p_{2})K^{A1}(p_{1})S^{AA}(p_{2},-p_{1})K^{A1}(p_{2})=K^{A1}(p_{2})S^{AA}(p_{1},-p_{2})K^{A1}(p_{1})S^{AA}(p_{1},p_{2})
\end{equation*}
one finds
that\begin{equation}
\frac{k_{2}}{k_{1}}=\frac{x_{B}+x^{+}}{x_{B}-x^{-}}\frac{\tilde{\eta}}{\eta}.\label{k2_k1_ratio}
\end{equation}
The calculations are done as follows. First one must consider the matrix element
\begin{equation}
\left\langle e_{1}\otimes e_{3}\right|\left(\mathrm{bYBE}\right)\left|e_{1}\otimes e_{3}\right\rangle,
\end{equation}
where $\left\langle e_{1}\otimes e_{3}\right|$ and $\left|e_{1}\otimes e_{3}\right\rangle$ are the orthonormal vectors (\ref{vect})
from the superspaces $\mathcal{V}^{1}\left(-p_{1},\zeta\right)\otimes\mathcal{V}^{1}\left(-p_{2},\zeta\mathrm{e}^{-i p_{1}}\right)$ 
and $\mathcal{V}^{1}\left(p_{1},\zeta\right)\otimes\mathcal{V}^{1}\left(p_{2},\zeta\mathrm{e}^{i p_{1}}\right)$ respectively.
This matrix element leads to the equation
\begin{eqnarray}
\left(k_{1}\left(p_{2}\right)\tilde{\eta}_{2}-k_{2}\left(p_{2}\right)\eta_{2}\right)\left(k_{2}\left(p_{1}\right)\eta_{1}\, x_{1}^{-}+k_{1}\left(p_{1}\right)\tilde{\eta}_{1}\,x_{1}^{+}\right) \qquad & & \notag \\
-\left(k_{1}\left(p_{1}\right)\tilde{\eta}_{1}-k_{2}\left(p_{1}\right)\eta_{1}\right)\left(k_{2}\left(p_{2}\right)\eta_{2}\, x_{2}^{-}+k_{1}\left(p_{2}\right)\tilde{\eta}_{2}\,x_{2}^{+}\right) & = & 0,
\end{eqnarray}
which may be solved by separating variables and setting
\begin{equation}
\frac{k_{2}\,\eta\,x^{-}+k_{1}\,\tilde{\eta}\,x^{+}}{k_{1}\,\tilde{\eta}-k_{2}\,\eta}=x_{B}.\label{k1k2xb}
\end{equation}
The solution of (\ref{k1k2xb}) gives the required equation (\ref{k2_k1_ratio}).
One must then show that the parameter $x_{B}$ is indeed the spectral parameter of the boundary state.
This may be achieved by considering the matrix element
\begin{equation}
\left\langle e_{1}\otimes e_{2}\right|\left(\mathrm{bYBE}\right)\left|e_{3}\otimes e_{4}\right\rangle,
\end{equation}
which may be set to zero only by requiring the parameter $x_{B}$ to satisfy the mass-shell condition (\ref{shortening_b}).

\subsection*{$K$-matrix $K^{Ba}$}

We define the $K$-matrix $K^{Ba}$ for reflection of a two-particle
bound state $B$ from the fundamental boundary state $a$ as a third-order
differential operator which acts as
\begin{equation*}
K^{Ba}\left(p,q\right):\quad\mathcal{V}^{2}\left(p,\zeta\right)\otimes\mathcal{V}^{1}\left(q,\zeta {\rm e}^{ip}\right)
\rightarrow\mathcal{V}^{2}\left(-p,\zeta\right)\otimes\mathcal{V}^{1}\left(q,\zeta {\rm e}^{-ip}\right),
\end{equation*}
corresponding to the tilded factor of the complete reflection $K$-matrix
\begin{equation*}
\tilde{K}:\:\boxslash\negthickspace\negthinspace\boxslash\otimes\boxslash\rightarrow\boxslash\negthickspace\negthinspace\boxslash\otimes\boxslash.
\end{equation*}
The $K$-matrix $K^{Ba}$ is given by the following differential operator
\begin{equation}
K^{Ba} = \sum_{i=1}^{19}k_{i}\,\Lambda_{i},
\end{equation}
where the  $\Lambda_{i}$ may be easily obtained using the method described in section 3.2 of \cite{Arutyunov1}. The symmetry
algebra fixes the values of the coefficients $k_{i}$ uniquely up
to a constant; the full expression may be found in the appendix.

\subsection*{$K$-matrix $K^{B1}$}

We define the $K$-matrix $K^{B1}$ for reflection of a two-particle bound
state $B$ from the singlet boundary state $1$ as a second-order
differential operator which acts as
\begin{equation*}
K^{B1}\left(p,q\right):\quad\mathcal{V}^{2}\left(p,\zeta\right)\otimes1\rightarrow\mathcal{V}^{2}\left(-p,\zeta\right)\otimes1,
\end{equation*}
corresponding to the untilded factor of the complete reflection $K$-matrix
\begin{equation*}
K:\:\boxslash\negthickspace\negthinspace\boxslash\otimes1\rightarrow\boxslash\negthickspace\negthinspace\boxslash\otimes1.
\end{equation*}
The $K$-matrix $K^{B1}$ is given by the sum of diagonal differential operators
\begin{equation}
K^{B1} = \sum_{i=1}^{3}k_{i}\,\Lambda_{i},
\end{equation}
where
\begin{equation}
\Lambda_{1} = \omega_{b}^{1}\omega_{a}^{1}\frac{\partial^{2}}{\partial\omega_{b}^{1}\partial\omega_{a}^{1}},\qquad
\Lambda_{2} = \omega_{a}^{1}\theta_{\alpha}^{1}\frac{\partial^{2}}{\partial\omega_{a}^{1}\partial\theta_{\alpha}^{1}},\qquad
\Lambda_{3} = \theta_{\beta}^{1}\theta_{\alpha}^{1}\frac{\partial^{2}}{\partial\theta_{\beta}^{1}\partial\theta_{\alpha}^{1}}.
\end{equation}
Once again, the symmetry algebra alone is not enough to fix coefficients
$k_{1}$, $k_{2}$ and $k_{3}$. We shall be using the bYBE
\begin{equation*}
S^{AB}(p_{1},p_{2})K^{A1}(p_{1})S^{AB}(p_{1},-p_{2})K^{B1}(p_{2})=K^{B1}(p_{2})S^{AB}(p_{1},-p_{2})K^{A1}(p_{1})S^{AB}(p_{1},p_{2}).
\end{equation*}
First we consider the matrix element
\begin{equation}
\left\langle e_{3}\otimes e_{1,1}\right|\left(\mathrm{bYBE}\right)\left|e_{1}\otimes e_{1,3}\right\rangle.
\end{equation}
Setting it to zero, we get a relation very similar to (\ref{k2_k1_ratio})
\begin{equation}
\frac{k_{2}}{k_{1}}=\frac{x_{B}+y^{+}}{x_{B}-y^{-}}\frac{\tilde{\eta}}{\eta}.
\end{equation}
The second constraint is acquired by considering the matrix element
\begin{equation}
\left\langle e_{3}\otimes e_{3,4}\right|\left(\mathrm{bYBE}\right)\left|e_{1}\otimes e_{2,3}\right\rangle,
\end{equation}
which gives the ratio
\begin{equation}
\frac{k_{3}}{k_{2}}=\frac{1-x_{B}y^{+}}{1+x_{B}y^{-}}\frac{\tilde{\eta}}{\eta}.
\end{equation}
Using the convenient normalization $k_{1}=1$, we obtain
\begin{eqnarray}
k_{1} & = & 1,\nonumber \\
k_{2} & = & \frac{x_{B}+y^{+}}{x_{B}-y^{-}}\frac{\tilde{\eta}}{\eta},\nonumber \\
k_{3} & = & \frac{(x_{B}+y^{+})(1-x_{B}y^{+})}{(x_{B}-y^{-})(1+x_{B}y^{-})}\frac{\tilde{\eta}^{2}}{\eta^{2}}.\label{KB1_coefs}
\end{eqnarray}

The reflection matrix $K^{B1}$ naturally extends to the reflection of any
bound state $K^{M1}$ with $M\geq2$ as there are always only three diagonal reflection coefficients with the following differential operators
\begin{eqnarray}
\Lambda_{1} & = & \omega_{a_{M}}^{1}\cdot\cdot\cdot\omega_{a_{1}}^{1}\frac{\partial^{M}}{\partial\omega_{a_{M}}^{1}\cdot\cdot\cdot\partial\omega_{a_{1}}^{1}},\nonumber \\
\Lambda_{2} & = & \omega_{a_{M-1}}^{1}\cdot\cdot\cdot\omega_{a_{1}}^{1}\theta_{\alpha}^{1}\frac{\partial^{M}}{\partial\omega_{a_{M-1}}^{1}\cdot\cdot\cdot\partial\omega_{a_{1}}^{1}\partial\theta_{\alpha}^{1}},\nonumber \\
\Lambda_{3} & = & \omega_{a_{M-2}}^{1}\cdot\cdot\cdot\omega_{a_{1}}^{1}\theta_{\beta}^{1}\theta_{\alpha}^{1}\frac{\partial^{M}}{\partial\omega_{a_{M-2}}^{1}\cdot\cdot\cdot\partial\omega_{a_{1}}^{1}\partial\theta_{\beta}^{1}\partial\theta_{\alpha}^{1}},
\end{eqnarray}
and the invariance under bYBE leads precisely to the same reflection coefficients as in (\ref{KB1_coefs})
subject to the mass shell conditions (\ref{shortening}) and
(\ref{shortening_b}).

\subsection*{$K$-matrix $K^{Ab}$}

We define the $K$-matrix $K^{Ab}$ for reflection of a fundamental bulk
state $A$ from a two-particle bound state $b$ on the boundary as
a third-order differential operator which acts as
\begin{equation*}
K^{Ab}\left(p,q\right):\quad\mathcal{V}^{1}\left(p,\zeta\right)\otimes\mathcal{V}^{2}\left(q,\zeta {\rm e}^{ip}\right)
\rightarrow\mathcal{V}^{1}\left(-p,\zeta\right)\otimes\mathcal{V}^{2}\left(q,\zeta {\rm e}^{-ip}\right).
\end{equation*}
corresponding to the tilded factor of the complete reflection $K$-matrix
\begin{equation*}
\tilde{K}:\:\boxslash\otimes\boxslash\negthinspace\boxslash\rightarrow\boxslash\otimes\boxslash\negthickspace\negthinspace\boxslash.
\end{equation*}
The $K$-matrix $K^{Ab}$ is given by the differential operator
\begin{equation}
K^{Ab} = \sum_{i=1}^{19}k_{i}\,\Lambda_{i},
\end{equation}
where the $\Lambda_{i}$ are as
in section 6.1.1 of \cite{Arutyunov1}. The symmetry algebra fixes
the values of the coefficients $k_{i}$ uniquely up to a constant; again the full
 expression may be found in the appendix.

\subsection*{$K$-matrix $K^{Bb}$}

We define the $K$-matrix $K^{Bb}$ for reflection of a bulk two-magnon bound
state $B$ from a two-magnon bound state $b$ on the boundary as a
fourth-order differential operator which acts as
\begin{equation*}
K^{Bb}\left(p,q\right):\quad\mathcal{V}^{2}\left(p,\zeta\right)\otimes\mathcal{V}^{2}\left(q,\zeta {\rm e}^{ip}\right)
\rightarrow\mathcal{V}^{2}\left(-p,\zeta\right)\otimes\mathcal{V}^{2}\left(q,\zeta {\rm e}^{-ip}\right).
\end{equation*}
corresponding to the tilded factor of the complete reflection $K$-matrix
\begin{equation*}
\tilde{K}:\:\boxslash\negthickspace\negthinspace\boxslash\otimes\boxslash\negthickspace\negthinspace\boxslash\rightarrow\boxslash\negthickspace\negthinspace\boxslash\otimes\boxslash\negthickspace\negthinspace\boxslash.
\end{equation*}
As a differential operator the $K$-matrix is
\begin{equation}
K^{Ab} = \sum_{i=1}^{48}k_{i}\,\Lambda_{i},
\end{equation}
where the $\Lambda_{i}$ are as given
in section 6.2.1 of \cite{Arutyunov1}. By choosing the constraints
to be $k_{1}=1$ and $k_{45},...,k_{48}=0$, in the same fashion as
we did for the bulk case, we were able to find the expressions for
$k_{2},...,k_{44}$ using the symmetry algebra alone. The values of the
coefficients $k_{i}$ may be found in the appendix.

\section{Conclusions and Outlook}

In this paper we have used the superspace formalism introduced in
\cite{Arutyunov1} to obtain the reflection matrices of magnon bound
states from the $Z=0$ D7 brane \cite{Young1}.
The matter fields in the bulk transform in representations of
$\mathfrak{psu}\left(2|2\right)\times\widetilde{\mathfrak{psu}}\left(2|2\right)$,
while on the boundary the residual Lie symmetry, preserved by the reflection of bulk excitations, is $\mathfrak{su}\left(2\right)\times\mathfrak{su}\left(2\right)\times\widetilde{\mathfrak{psu}}\left(2|2\right)$.
The reflection matrix factors as a tensor product for tilded and
untilded factors. The reflection of fundamental magnons was rigorously
worked out in \cite{Young1}.

Here we have calculated the scattering matrices $K^{Ba}$, $K^{Ab}$
and $K^{Bb}$ describing the reflections of two-magnon bound states
in the tilded factor. We found that $K^{Ba}$ and $K^{Ab}$ may be
determined up to an overall factor by the symmetry algebra alone, just as for the bulk $S^{BA}$
of \cite{Arutyunov1}.
However, as for the bulk $S^{BB}$, we found that $K^{Bb}$ is not determined
uniquely by the symmetry algebra, but must be constrained by the bYBE.
Alternatively, it turns out that we may set the coefficients
$k_{45},...,k_{48}$ to zero and then use the symmetry algebra alone.

Further, we have calculated the reflection matrix $K^{B1}$ describing
the reflection in the untilded factor of the two-magnon bound state in
the bulk from a singlet boundary state. This may naturally be
generalized to any $K^{M1}$ with $M\geq2$, as these have only
three reflection coefficients, which may be fixed
up to an overall scalar factor by requiring that the bYBE be satisfied.
This is in  agreement with the results of \cite{Young1}
for the reflection of fundamental states.

An important question is of the physicality of the boundary bound states.
The bound state $K$-matrices we have constructed here have a pole (and zero) structure
very similar to the bulk case, so that the pole $x^{+}=x_{B}$ appearing throughout the $K$-matrices
signals the presence of higher order multi-magnon bound states.
As was noted in \cite{Maldacena2}, an open string ending on a giant magnon
is expected to have a tower of multi-magnon bound states
on the boundary for any value of the coupling constant $g$. The tilded factor of the $Z=0$ D7 brane
shares the same symmetry as the $Z=0$ giant magnon; thus we expect multi-magnon
bound states to be living on the tilded factor of our boundary \cite{Ahn2}. But this is not the case for the untilded factor.
We do not expect the pole in the reflection matrix $K^{M1}$ to correspond to a physical bound state
for any $M$, since the values of $\Delta-J_{56}$ on the boundary fields
(and thereby states) \cite{Young1} are inconsistent
with the pole residues, at least for the low-lying states.
However, the issue of the physicality of putative bound states cannot be fully resolved without an analysis of
the boundary on-shell (Landau) diagrams
\cite{Palla1, PDorey1, PDorey2}, which is beyond the scope of the present work.

The problem remains of understanding the structure of the scattering matrices we have found.
One would expect that, where the bYBE was needed in addition to the Lie superalgebra,
the underlying Yangian symmetry should supply the deficit, as in the bulk case \cite{deLeeuw1,Arutyunov6}.
In the boundary case we would expect to see some form of generalized twisted Yangian \cite{MacKay2} as the boundary
remnant of the bulk Yangian symmetry, and that this would, for example, be sufficient to fix $K^{B1}$ up to an overall factor,
as happens in the maximal giant graviton (D3) case \cite{Ahn1}. Similarly for $K^{Bb}$ we would hope that the Yangian
symmetry would explain the curious zeros in $k_{45}, ... , k_{48}$, and organize the coefficients of $K^{Bb}$ more sensibly.
This is the subject of current investigations.

\vskip 0.2in
\textit{\bf Acknowledgments:} We thank Charles Young,
Niklas Beisert and Alessandro Torrielli for helpful discussions, and the UK EPSRC
for funding under grant EP/H000054/1.

\newpage

\appendix

\section{Appendices}

\subsection{Coefficients of $S$-matrices}

We use the convenient notation of the bulk spectral parameters, where
$x_{1}^{\pm}$, $x_{2}^{\pm}$ are the parameters of the fundamental
states with momentum $p_{1}$ and $p_{2}$, while $y_{1}^{\pm}$,
$y_{2}^{\pm}$ are the parameters of the two-particle bound states
respectively with momentum $p_{1}$ and $p_{2}$. The parameters
$\eta_{i}$ are given by
\begin{equation}
\eta_{1}=\eta\left(p_{1}\right),\qquad\eta_{2}={\rm e}^{\frac{i}{2}p_{1}}\eta\left(p_{2}\right),\qquad\tilde{\eta}_{1}=
{\rm e}^{\frac{i}{2}p_{2}}\eta\left(p_{1}\right),\qquad\tilde{\eta}_{2}=\eta\left(p_{2}\right),
\end{equation}
where $\eta\left(p_{i}\right)={\rm e}^{i\xi_{0}}{\rm e}^{\frac{i}{4}p_{i}}\sqrt{i\left(x_{i}^{-}-x_{i}^{+}\right)}$
for the fundamental states and $x_{i}^{\pm}$ must be changed into $y_{i}^{\pm}$
for the bound states.
The rapidity parameters $u_{i}$ used in the expressions for the coefficients
of the $S$-matrix $S^{BB}$ are expressed in terms of $y_{i}^{\pm}$ as
follows
\begin{equation}
u_{i}=\frac{1}{2}\left(y_{i}^{+}+\frac{1}{y_{i}^{+}}+y_{i}^{-}+\frac{1}{y_{i}^{-}}\right).
\end{equation}

\begin{flushleft}
\subsubsection*{The $S$-matrix $S^{AA}$}

The $S$-matrix $S^{AA}$ has the following components $a_{i}$:

$
\begin{aligned}
a_{1} & = 1,\\
a_{2} & = 2\frac{x_{2}^{+}\left(x_{1}^{-}-x_{2}^{-}\right)\left(-1+x_{1}^{+}x_{2}^{-}\right)}{x_{2}^{-}\left(x_{1}^{-}-x_{2}^{+}\right)\left(-1+x_{1}^{+}x_{2}^{+}\right)}-1,\\
a_{3} & = -\frac{\left(x_{2}^{-}-x_{1}^{+}\right)}{\left(x_{1}^{-}-x_{2}^{+}\right)}\frac{\tilde{\eta}_{1}\tilde{\eta_{2}}}{\eta_{1}\eta_{2}},\\
a_{4} & = \left(\frac{x_{2}^{-}-x_{1}^{+}}{x_{1}^{-}-x_{2}^{+}}+2\frac{x_{1}^{+}\left(x_{1}^{-}-x_{2}^{-}\right)\left(-1+x_{1}^{-}x_{2}^{+}\right)}{x_{1}^{-}\left(x_{1}^{-}-x_{2}^{+}\right)\left(-1+x_{1}^{+}x_{2}^{+}\right)}\right)\frac{\tilde{\eta}_{1}\tilde{\eta}_{2}}{\eta_{1}\eta_{2}},\\
a_{5} & = \frac{\left(x_{1}^{+}-x_{2}^{+}\right)}{\left(x_{1}^{-}-x_{2}^{+}\right)}\frac{\tilde{\eta}_{2}}{\eta_{1}},\\
a_{6} & = \frac{\left(x_{1}^{-}-x_{2}^{-}\right)}{\left(x_{1}^{-}-x_{2}^{+}\right)}\frac{\tilde{\eta}_{1}}{\eta_{1}},\\
a_{7} & = -\frac{i\zeta x_{1}^{+}\left(x_{1}^{-}-x_{2}^{-}\right)\left(x_{1}^{-}-x_{1}^{+}\right)\left(x_{2}^{-}-x_{2}^{+}\right)}{x_{1}^{-}x_{2}^{-}\left(x_{1}^{-}-x_{2}^{+}\right)\left(-1+x_{1}^{+}x_{2}^{+}\right)\eta_{1}\eta_{2}},\\
a_{8} & = \frac{i\left(x_{1}^{-}-x_{2}^{-}\right)\tilde{\eta}_{1}\tilde{\eta}_{2}}{\zeta\left(x_{1}^{-}-x_{2}^{+}\right)\left(-1+x_{1}^{+}x_{2}^{+}\right)},\\
a_{9} & = \frac{\left(x_{1}^{-}-x_{1}^{+}\right)}{\left(x_{1}^{-}-x_{2}^{+}\right)}\frac{\tilde{\eta}_{2}}{\eta_{1}},\\
a_{10} & = \frac{\left(x_{2}^{-}-x_{2}^{+}\right)}{\left(x_{1}^{-}-x_{2}^{+}\right)}\frac{\tilde{\eta}_{1}}{\eta_{2}}.
\end{aligned}
$

\noindent These coefficients are in agreement with the ones found in \cite{Beisert2}
up to a relative  sign corresponding to the eigenvalue of the graded permutation operator acting on the antisymmetric states (see footnote on page 11) and an overall factor
\begin{equation*}
 N_{0}^{AA}=\frac{x_{1}^{-}-x_{2}^{+}}{x_{2}^{-}-x_{1}^{+}}.
\end{equation*}
This corresponds to the normalization $a_{3}=1$, which we might refer to as the `physical
normalization' since it removes the pole in the $S$-matrix element $a_{3}$, which would produce a state
symmetric in fermionic indices state and therefore cannot yield a bound state. Rather it is the pole at $x_{2}^{-}-x_{1}^{+}$ which is responsible for the creation of bound states, although this pole is typically hidden in the normalization $a_{1} = 1$ used in calculations.

\subsubsection*{The $S$-matrix $S^{AB}$}

The $S$-matrix $S^{AB}$ has the following components $a_{i}$:

$
\begin{aligned}
a_{1} & = 1,\\
a_{2} & = -\frac{1}{2}+\frac{3y_{2}^{+}\left(x_{1}^{-}-y_{2}^{-}\right)\left(-1+x_{1}^{+}y_{2}^{-}\right)}{2y_{2}^{-}\left(x_{1}^{-}-y_{2}^{+}\right)\left(-1+x_{1}^{+}y_{2}^{+}\right)},\\
a_{3} & = \frac{\left(x_{1}^{+}-y_{2}^{+}\right)}{\left(x_{1}^{-}-y_{2}^{+}\right)}\frac{\tilde{\eta}_{2}}{\eta_{2}},\\
a_{4} & = \frac{y_{2}^{+}\left(x_{1}^{+}-y_{2}^{-}\right)\left(-1+x_{1}^{+}y_{2}^{-}\right)}{y_{2}^{-}\left(x_{1}^{-}-y_{2}^{+}\right)\left(-1+x_{1}^{+}y_{2}^{+}\right)}\frac{\tilde{\eta}_{2}}{\eta_{2}},\\
a_{5} & = \frac{\left(x_{1}^{-}-y_{2}^{-}\right)}{\left(x_{1}^{-}-y_{2}^{+}\right)}\frac{\tilde{\eta}_{1}}{\eta_{1}},\\
a_{6} & = \frac{x_{1}^{+}\left(x_{1}^{+}y_{2}^{-}+x_{1}^{+}y_{2}^{+}-2y_{2}^{-}y_{2}^{+}-2x_{1}^{-}x_{1}^{+}y_{2}^{-}y_{2}^{+}+x_{1}^{-}\left(y_{2}^{-}\right)^{2}y_{2}^{+}+x_{1}^{-}y_{2}^{-}\left(y_{2}^{+}\right)^{2}\right)\tilde{\eta}_{2}^{2}}{2x_{1}^{-}y_{2}^{-}\left(x_{1}^{-}-y_{2}^{+}\right)\left(1-x_{1}^{+}y_{2}^{+}\right)\eta_{2}^{2}},\\
a_{7} & = -\frac{\left(x_{1}^{+}-y_{2}^{-}\right)}{\left(x_{1}^{-}-y_{2}^{+}\right)}\frac{\tilde{\eta}_{1}\tilde{\eta}_{2}}{\eta_{1}\eta_{2}},\\
a_{8} & = -\left(\frac{2x_{1}^{+}\left(x_{1}^{-}-y_{2}^{-}\right)\left(-1+x_{1}^{-}y_{2}^{+}\right)}{x_{1}^{-}\left(x_{1}^{-}-y_{2}^{+}\right)\left(-1+x_{1}^{+}y_{2}^{+}\right)}-\frac{x_{1}^{+}-y_{2}^{-}}{x_{1}^{-}-y_{2}^{+}}\right)\frac{\tilde{\eta}_{1}\tilde{\eta}_{2}}{\eta_{1}\eta_{2}},\\
a_{9} & = \frac{x_{1}^{+}\left(x_{1}^{+}-y_{2}^{-}\right)\left(-1+x_{1}^{-}y_{2}^{+}\right)}{x_{1}^{-}\left(x_{1}^{-}-y_{2}^{+}\right)\left(-1+x_{1}^{+}y_{2}^{+}\right)}\frac{\tilde{\eta}_{1}\tilde{\eta}_{2}^{2}}{\eta_{1}\eta_{2}^{2}},\\
a_{10} & = -\frac{i\zeta\left(x_{1}^{-}-x_{1}^{+}\right)x_{1}^{+}\left(y_{2}^{-}-y_{2}^{+}\right)^{2}y_{2}^{+}}{2x_{1}^{-}y_{2}^{-}\left(x_{1}^{-}-y_{2}^{+}\right)\left(-1+x_{1}^{+}y_{2}^{+}\right)\eta_{2}^{2}},\\
a_{11} & = \frac{i\left(x_{1}^{-}-x_{1}^{+}\right)\tilde{\eta}_{2}^{2}}{2\zeta\left(x_{1}^{-}-y_{2}^{+}\right)\left(-1+x_{1}^{+}y_{2}^{+}\right)},\\
a_{12} & = \frac{i\zeta x_{1}^{+}y_{2}^{+}\left(x_{1}^{-}-x_{1}^{+}\right)\left(x_{1}^{-}-y_{2}^{-}\right)\left(y_{2}^{-}-y_{2}^{+}\right)}{\sqrt{2}x_{1}^{-}y_{2}^{-}\left(-x_{1}^{-}+y_{2}^{+}\right)\left(-1+x_{1}^{+}y_{2}^{+}\right)\eta_{1}\eta_{2}},\\
\end{aligned}
$

$
\begin{aligned}
a_{13} & = -\frac{i\left(x_{1}^{-}-y_{2}^{-}\right)\tilde{\eta}_{1}\tilde{\eta}_{2}}{\sqrt{2}\zeta\left(x_{1}^{-}-y_{2}^{+}\right)\left(-1+x_{1}^{+}y_{2}^{+}\right)},\\
a_{14} & = \frac{\left(x_{1}^{-}-x_{1}^{+}\right)x_{1}^{+}\left(-1+x_{1}^{-}y_{2}^{+}\right)}{\sqrt{2}x_{1}^{-}\left(x_{1}^{-}-y_{2}^{+}\right)\left(-1+x_{1}^{+}y_{2}^{+}\right)}\frac{\tilde{\eta}_{2}^{2}}{\eta_{1}\eta_{2}},\\
a_{15} & = \frac{x_{1}^{+}\left(-y_{2}^{-}+y_{2}^{+}\right)\left(-1+x_{1}^{-}y_{2}^{+}\right)}{\sqrt{2}x_{1}^{-}\left(x_{1}^{-}-y_{2}^{+}\right)\left(-1+x_{1}^{+}y_{2}^{+}\right)}\frac{\tilde{\eta}_{1}\tilde{\eta}_{2}}{\eta_{2}^{2}},\\
a_{16} & = \frac{i\left(x_{1}^{+}-y_{2}^{-}\right)}{\sqrt{2}\zeta\left(x_{1}^{-}-y_{2}^{+}\right)\left(-1+x_{1}^{+}y_{2}^{+}\right)}\frac{\tilde{\eta}_{1}\tilde{\eta}_{2}^{2}}{\eta_{2}},\\
a_{17} & = \frac{i\zeta x_{1}^{+}y_{2}^{+}\left(x_{1}^{-}-x_{1}^{+}\right)\left(x_{1}^{+}-y_{2}^{-}\right)\left(y_{2}^{-}-y_{2}^{+}\right)}{\sqrt{2}x_{1}^{-}y_{2}^{-}\left(-x_{1}^{-}+y_{2}^{+}\right)\left(-1+x_{1}^{+}y_{2}^{+}\right)}\frac{\tilde{\eta}_{2}}{\eta_{1}\eta_{2}^{2}},\\
a_{18} & = \frac{\left(x_{1}^{-}-x_{1}^{+}\right)}{\sqrt{2}\left(x_{1}^{-}-y_{2}^{+}\right)}\frac{\tilde{\eta}_{2}}{\eta_{1}},\\
a_{19} & = \frac{\left(y_{2}^{-}-y_{2}^{+}\right)}{\sqrt{2}\left(x_{1}^{-}-y_{2}^{+}\right)}\frac{\tilde{\eta}_{1}}{\eta_{2}}.
\end{aligned}
$

\subsubsection*{The $S$-matrix $S^{BB}$}

The $S$-matrix $S^{BB}$ has the following components $a_{i}$:

$
\begin{aligned}
a_{1} & = 1,\\
a_{2} & = \frac{\left(y_{2}^{-}-y_{1}^{+}\right)\left(-1+y_{2}^{-}y_{1}^{+}\right)y_{2}^{+}}{\left(-1+y_{1}^{-}y_{2}^{-}\right)y_{1}^{+}\left(y_{1}^{-}-y_{2}^{+}\right)}\\
 & \quad\times\frac{\left(y_{1}^{-}\left(2+y_{2}^{-}\left(y_{1}^{-}-3y_{1}^{+}\right)\right)y_{1}^{+}+\left(y_{1}^{+}+y_{1}^{-}\left(-3+2y_{2}^{-}y_{1}^{+}\right)\right)y_{2}^{+}\right)}{\left(-3y_{2}^{-}y_{1}^{+}+\left(y_{1}^{+}+y_{2}^{-}\left(2+y_{1}^{+}\left(y_{2}^{-}+2y_{1}^{+}\right)\right)\right)y_{2}^{+}-3y_{2}^{-}y_{1}^{+}\left(y_{2}^{+}\right)^{2}\right)},\\
a_{3} & = \frac{1}{2y_{2}^{-}\left(-1+y_{1}^{-}y_{2}^{-}\right)\left(y_{1}^{+}\right)^{2}\left(y_{1}^{-}-y_{2}^{+}\right)y_{2}^{+}\left(u_{1}-u_{2}-\frac{2i}{g}\right)}\\
 & \quad\times\Biggl(\left(y_{1}^{-}\right)^{2}\left(y_{2}^{-}\right)^{2}\left(y_{1}^{+}\right)^{2}+\left(y_{1}^{+}\right)^{2}\left(y_{2}^{-}\left(2+\left(y_{2}^{-}-y_{2}^{+}\right)^{2}\right)-y_{2}^{+}\right)y_{2}^{+}\\
 & \quad\qquad+y_{1}^{-}y_{2}^{-}\Bigl(2y_{2}^{-}\left(y_{1}^{+}\right)^{4}y_{2}^{+}+2\left(y_{2}^{+}\right)^{2}-y_{1}^{+}y_{2}^{+}\left(5+\left(y_{2}^{-}\right)^{2}+y_{2}^{-}y_{2}^{+}+2\left(y_{2}^{+}\right)^{2}\right)\\
 & \quad\hspace{2.3cm}+\left(y_{1}^{+}\right)^{2}\left(2+\left(y_{2}^{-}\right)^{2}-y_{2}^{-}\left(-1+\left(y_{2}^{-}\right)^{2}\right)y_{2}^{+}+3\left(2+\left(y_{2}^{-}\right)^{2}\right)\left(y_{2}^{+}\right)^{2}\right)\\
 & \quad\hspace{2.3cm}-\left(y_{1}^{+}\right)^{3}\left(3y_{2}^{+}+y_{2}^{-}\left(2+y_{2}^{+}\left(y_{2}^{-}+3y_{2}^{+}\right)\right)\right)\Bigr)\Biggr),\\
a_{4} & = -\frac{\left(y_{2}^{-}-y_{1}^{+}\right)}{\left(y_{1}^{-}-y_{2}^{+}\right)}\frac{\tilde{\eta}_{1}\tilde{\eta}_{2}}{\eta_{1}\eta_{2}},\\
a_{5} & = -\left(\frac{2\left(y_{1}^{-}-y_{2}^{-}\right)\left(-y_{1}^{+}+y_{2}^{+}\right)\left(-1+y_{1}^{-}y_{2}^{+}\right)}{y_{1}^{-}\left(y_{1}^{-}-y_{2}^{+}\right)y_{2}^{+}\left(u_{1}-u_{2}-\frac{2i}{g}\right)}-\frac{\left(y_{2}^{-}-y_{1}^{+}\right)}{\left(y_{1}^{-}-y_{2}^{+}\right)}\right)\frac{\tilde{\eta}_{1}\tilde{\eta}_{2}}{\eta_{1}\eta_{2}},\\
\end{aligned}
$

$
\begin{aligned}
a_{6} & = -\frac{\left(y_{2}^{-}-y_{1}^{+}\right)}{\left(y_{1}^{-}-y_{2}^{+}\right)}\frac{\left(u_{1}-u_{2}+\frac{2i}{g}\right)}{\left(u_{1}-u_{2}-\frac{2i}{g}\right)}\frac{\tilde{\eta}_{1}\tilde{\eta}_{2}}{\eta_{1}\eta_{2}},\\
a_{7} & = -\frac{\left(y_{2}^{-}-y_{1}^{+}\right)}{2y_{2}^{-}\left(-1+y_{1}^{-}y_{2}^{-}\right)y_{1}^{+}\left(y_{1}^{-}-y_{2}^{+}\right)y_{2}^{+}\left(u_{1}-u_{2}-\frac{2i}{g}\right)}\frac{\tilde{\eta}_{1}\tilde{\eta}_{2}}{\eta_{1}\eta_{2}}\\
 & \quad\times\Biggl(2y_{2}^{-}y_{1}^{+}y_{2}^{+}\left(u_{1}-u_{2}+\frac{2i}{g}\right)+y_{2}^{-}\Bigl(-2\left(2+y_{1}^{-}\left(y_{2}^{-}-2y_{2}^{+}\right)\right)y_{2}^{+}\\
 & \quad\quad\qquad+2y_{2}^{-}\left(y_{1}^{+}\right)^{2}\left(-2+y_{1}^{-}y_{2}^{+}\right)+y_{1}^{+}\left(4+y_{1}^{-}\left(y_{2}^{-}-3y_{2}^{+}\right)\right)\left(1+y_{2}^{-}y_{2}^{+}\right)\Bigr)\Biggr),\\
a_{8} & = \frac{\left(y_{2}^{-}-y_{1}^{+}\right)\left(y_{1}^{-}y_{2}^{-}-2y_{1}^{-}y_{1}^{+}+y_{2}^{-}y_{1}^{+}\right)\left(-1+y_{1}^{-}y_{2}^{+}\right)}{2\left(y_{1}^{-}\right)^{2}y_{2}^{+}\left(-y_{1}^{-}+y_{2}^{+}\right)\left(-1+y_{1}^{+}y_{2}^{+}\right)\left(u_{1}-u_{2}-\frac{2i}{g}\right)}\frac{\tilde{\eta}_{1}^{2}\tilde{\eta}_{2}^{2}}{\eta_{1}^{2}\eta_{2}^{2}}\\
 & \quad-\frac{\left(y_{2}^{-}-y_{1}^{+}\right)y_{1}^{+}\left(y_{1}^{-}-2y_{2}^{-}+y_{1}^{+}\right)\left(-1+y_{1}^{-}y_{2}^{+}\right)}{2y_{1}^{-}\left(y_{1}^{-}-y_{2}^{+}\right)\left(-1+y_{1}^{+}y_{2}^{+}\right)\left(u_{1}-u_{2}-\frac{2i}{g}\right)}\frac{\tilde{\eta}_{1}^{2}\tilde{\eta}_{2}^{2}}{\eta_{1}^{2}\eta_{2}^{2}},\\
a_{9} & = \frac{\left(y_{1}^{+}-y_{2}^{+}\right)}{\left(y_{1}^{-}-y_{2}^{+}\right)}\frac{\tilde{\eta}_{2}}{\eta_{2}},\\
a_{10} & = \frac{\left(y_{1}^{-}-y_{2}^{-}\right)}{\left(y_{1}^{-}-y_{2}^{+}\right)}\frac{\tilde{\eta}_{1}}{\eta_{1}},\\
a_{11} & = -\frac{\left(y_{2}^{-}-y_{1}^{+}\right)\left(y_{1}^{+}-y_{2}^{+}\right)\left(-1+y_{2}^{-}y_{1}^{+}\right)}{y_{2}^{-}y_{1}^{+}\left(y_{1}^{-}-y_{2}^{+}\right)\left(u_{1}-u_{2}-\frac{2i}{g}\right)}\frac{\tilde{\eta}_{2}}{\eta_{2}},\\
a_{12} & = \frac{\left(-y_{1}^{-}+y_{2}^{-}\right)\left(y_{2}^{-}-y_{1}^{+}\right)\left(-1+y_{2}^{-}y_{1}^{+}\right)}{y_{2}^{-}y_{1}^{+}\left(y_{1}^{-}-y_{2}^{+}\right)\left(u_{1}-u_{2}-\frac{2i}{g}\right)}\frac{\tilde{\eta}_{1}}{\eta_{1}},\\
a_{13} & = \frac{\left(y_{1}^{+}-y_{2}^{+}\right)\tilde{\eta}_{2}^{2}}{2y_{1}^{-}y_{2}^{-}\left(-1+y_{1}^{-}y_{2}^{-}\right)y_{1}^{+}\left(y_{1}^{-}-y_{2}^{+}\right)y_{2}^{+}\left(u_{1}-u_{2}-\frac{2i}{g}\right)\eta_{2}^{2}},\\
 & \quad\times\Biggl(y_{1}^{+}\left(y_{2}^{-}y_{1}^{+}+\left(y_{1}^{+}+y_{2}^{-}\left(-2+\left(y_{2}^{-}\right)^{2}-2y_{2}^{-}y_{1}^{+}\right)\right)y_{2}^{+}+\left(y_{2}^{-}\right)^{2}\left(y_{2}^{+}\right)^{2}\right)\\
 & \quad\qquad+y_{1}^{-}y_{2}^{-}\Bigl(-\left(y_{2}^{-}\right)^{3}y_{1}^{+}y_{2}^{+}-\left(y_{1}^{+}\right)^{2}y_{2}^{+}+\left(y_{2}^{-}\right)^{2}\left(y_{1}^{+}+\left(-1+\left(y_{1}^{+}\right)^{2}\right)y_{2}^{+}\right)\\
 & \qquad\qquad+y_{2}^{-}\left(2\left(y_{1}^{+}\right)^{3}y_{2}^{+}-\left(y_{2}^{+}\right)^{2}-3\left(y_{1}^{+}\right)^{2}\left(1+\left(y_{2}^{+}\right)^{2}\right)+y_{1}^{+}y_{2}^{+}\left(5+\left(y_{2}^{+}\right)^{2}\right)\right)\Bigr)\Biggr),\\
a_{14} & = \left(\frac{\left(y_{1}^{-}-y_{2}^{-}\right)\left(y_{1}^{-}-2y_{2}^{-}+y_{1}^{+}\right)}{2\left(y_{1}^{-}-y_{2}^{+}\right)\left(u_{1}-u_{2}-\frac{2i}{g}\right)}+\frac{\left(y_{1}^{-}-y_{2}^{-}\right)\left(y_{1}^{-}\left(y_{2}^{-}-2y_{1}^{+}\right)+y_{2}^{-}y_{1}^{+}\right)}{2y_{1}^{-}y_{1}^{+}\left(y_{1}^{-}-y_{2}^{+}\right)y_{2}^{+}\left(u_{1}-u_{2}-\frac{2i}{g}\right)}\right)\frac{\tilde{\eta}_{1}^{2}}{\eta_{1}^{2}},\\
a_{15} & = -\frac{\left(y_{1}^{-}-y_{2}^{-}\right)\left(y_{2}^{-}-y_{1}^{+}\right)\left(-1+y_{1}^{-}y_{2}^{+}\right)}{y_{2}^{+}\left(\left(y_{1}^{-}\right)^{2}-y_{1}^{-}y_{2}^{+}\right)\left(u_{1}-u_{2}-\frac{2i}{g}\right)}\frac{\tilde{\eta}_{1}^{2}\tilde{\eta}_{2}}{\eta_{1}^{2}\eta_{2}},\\
a_{16} & = \frac{\left(y_{2}^{-}-y_{1}^{+}\right)\left(y_{1}^{+}-y_{2}^{+}\right)\left(-1+y_{1}^{-}y_{2}^{+}\right)}{y_{1}^{-}\left(-y_{1}^{-}y_{2}^{+}+\left(y_{2}^{+}\right)^{2}\right)\left(u_{1}-u_{2}-\frac{2i}{g}\right)}\frac{\tilde{\eta}_{1}\tilde{\eta}_{2}^{2}}{\eta_{1}\eta_{2}^{2}},\\
\end{aligned}
$

$
\begin{aligned}
a_{17} & = -\frac{\zeta^{2}\left(y_{1}^{-}-y_{2}^{-}\right)\left(y_{1}^{-}-y_{1}^{+}\right)^{2}\left(y_{2}^{-}-y_{1}^{+}\right)y_{1}^{+}\left(y_{2}^{-}-y_{2}^{+}\right)^{2}y_{2}^{+}}{2\left(y_{1}^{-}\right)^{2}\left(y_{2}^{-}\right)^{2}\left(y_{1}^{-}-y_{2}^{+}\right)\left(-1+y_{1}^{+}y_{2}^{+}\right)\left(u_{1}-u_{2}-\frac{2i}{g}\right)\eta_{1}^{2}\eta_{2}^{2}},\\
a_{18} & = -\frac{y_{1}^{-}y_{2}^{-}\left(y_{2}^{-}-y_{1}^{+}\right)\left(y_{1}^{+}-y_{2}^{+}\right)\tilde{\eta}_{1}^{2}\tilde{\eta}_{2}^{2}}{2\zeta^{2}\left(-1+y_{1}^{-}y_{2}^{-}\right)\left(y_{1}^{+}\right)^{2}\left(y_{1}^{-}-y_{2}^{+}\right)\left(y_{2}^{+}\right)^{2}\left(u_{1}-u_{2}-\frac{2i}{g}\right)},\\
a_{19} & = \frac{i\zeta\left(y_{1}^{-}-y_{1}^{+}\right)\left(y_{2}^{-}-y_{1}^{+}\right)\left(-1+y_{2}^{-}y_{1}^{+}\right)\left(y_{2}^{-}-y_{2}^{+}\right)\left(y_{1}^{+}-y_{2}^{+}\right)}{2y_{2}^{-}\left(-1+y_{1}^{-}y_{2}^{-}\right)y_{1}^{+}\left(y_{1}^{-}-y_{2}^{+}\right)\left(u_{1}-u_{2}-\frac{2i}{g}\right)\eta_{1}\eta_{2}},\\
a_{20} & = -\frac{iy_{1}^{-}\left(y_{2}^{-}-y_{1}^{+}\right)\left(-1+y_{2}^{-}y_{1}^{+}\right)\left(y_{1}^{+}-y_{2}^{+}\right)\tilde{\eta}_{1}\tilde{\eta}_{2}}{2\zeta\left(-1+y_{1}^{-}y_{2}^{-}\right)\left(y_{1}^{+}\right)^{2}\left(y_{1}^{-}-y_{2}^{+}\right)y_{2}^{+}\left(u_{1}-u_{2}-\frac{2i}{g}\right)},\\
a_{21} & = \frac{iy_{2}^{-}\left(y_{2}^{-}-y_{1}^{+}\right)\left(y_{1}^{+}-y_{2}^{+}\right)\left(-1+y_{1}^{-}y_{2}^{+}\right)}{2\zeta\left(-1+y_{1}^{-}y_{2}^{-}\right)y_{1}^{+}\left(y_{1}^{-}-y_{2}^{+}\right)\left(y_{2}^{+}\right)^{2}\left(u_{1}-u_{2}-\frac{2i}{g}\right)}\frac{\tilde{\eta}_{1}^{2}\tilde{\eta}_{2}^{2}}{\eta_{1}\eta_{2}},\\
a_{22} & = -\frac{i\zeta\left(y_{1}^{-}-y_{2}^{-}\right)\left(y_{1}^{-}-y_{1}^{+}\right)\left(y_{2}^{-}-y_{1}^{+}\right)y_{1}^{+}\left(y_{2}^{-}-y_{2}^{+}\right)\left(-1+y_{1}^{-}y_{2}^{+}\right)}{2\left(y_{1}^{-}\right)^{2}y_{2}^{-}\left(y_{1}^{-}-y_{2}^{+}\right)\left(-1+y_{1}^{+}y_{2}^{+}\right)\left(u_{1}-u_{2}-\frac{2i}{g}\right)}\frac{\tilde{\eta}_{1}\tilde{\eta}_{2}}{\eta_{1}^{2}\eta_{2}^{2}},\\
a_{23} & = \frac{i\zeta\left(y_{1}^{-}-y_{1}^{+}\right)\left(y_{2}^{-}-y_{2}^{+}\right)\left(y_{1}^{+}-y_{2}^{+}\right)^{2}\left(-1+y_{1}^{+}y_{2}^{+}\right)}{2\left(-1+y_{1}^{-}y_{2}^{-}\right)y_{1}^{+}y_{2}^{+}\left(-y_{1}^{-}+y_{2}^{+}\right)\left(u_{1}-u_{2}-\frac{2i}{g}\right)\eta_{1}\eta_{2}},\\
a_{24} & = \frac{iy_{1}^{-}y_{2}^{-}\left(y_{1}^{+}-y_{2}^{+}\right)^{2}\left(-1+y_{1}^{+}y_{2}^{+}\right)\tilde{\eta}_{1}\tilde{\eta}_{2}}{2\zeta\left(-1+y_{1}^{-}y_{2}^{-}\right)\left(y_{1}^{+}\right)^{2}\left(y_{1}^{-}-y_{2}^{+}\right)\left(y_{2}^{+}\right)^{2}\left(u_{1}-u_{2}-\frac{2i}{g}\right)},\\
a_{25} & = \frac{\left(y_{1}^{-}-y_{1}^{+}\right)^{2}\left(-1+y_{1}^{-}y_{2}^{+}\right)}{2y_{1}^{-}\left(y_{1}^{-}-y_{2}^{+}\right)y_{2}^{+}\left(u_{1}-u_{2}-\frac{2i}{g}\right)}\frac{\tilde{\eta}_{2}^{2}}{\eta_{1}^{2}},\\
a_{26} & = \frac{\left(y_{2}^{-}-y_{2}^{+}\right)^{2}\left(-1+y_{1}^{-}y_{2}^{+}\right)}{2y_{1}^{-}\left(y_{1}^{-}-y_{2}^{+}\right)y_{2}^{+}\left(u_{1}-u_{2}-\frac{2i}{g}\right)}\frac{\tilde{\eta}_{1}^{2}}{\eta_{2}^{2}},\\
a_{27} & = \frac{\left(y_{1}^{-}-y_{1}^{+}\right)\left(-y_{2}^{-}+y_{1}^{+}\right)\left(-1+y_{2}^{-}y_{1}^{+}\right)}{2y_{2}^{-}y_{1}^{+}\left(y_{1}^{-}-y_{2}^{+}\right)\left(u_{1}-u_{2}-\frac{2i}{g}\right)}\frac{\tilde{\eta}_{2}}{\eta_{1}},\\
a_{28} & = -\frac{\left(y_{2}^{-}-y_{1}^{+}\right)\left(-1+y_{2}^{-}y_{1}^{+}\right)\left(y_{2}^{-}-y_{2}^{+}\right)}{2y_{2}^{-}y_{1}^{+}\left(y_{1}^{-}-y_{2}^{+}\right)\left(u_{1}-u_{2}-\frac{2i}{g}\right)}\frac{\tilde{\eta}_{1}}{\eta_{2}},\\
a_{29} & = \frac{\left(y_{1}^{-}-y_{1}^{+}\right)\left(-y_{2}^{-}+y_{1}^{+}\right)\left(-1+y_{1}^{-}y_{2}^{+}\right)}{2y_{2}^{+}\left(\left(y_{1}^{-}\right)^{2}-y_{1}^{-}y_{2}^{+}\right)\left(u_{1}-u_{2}-\frac{2i}{g}\right)}\frac{\tilde{\eta}_{1}\tilde{\eta}_{2}^{2}}{\eta_{1}^{2}\eta_{2}},\\
a_{30} & = \frac{\left(y_{2}^{-}-y_{1}^{+}\right)\left(y_{2}^{-}-y_{2}^{+}\right)\left(-1+y_{1}^{-}y_{2}^{+}\right)}{2y_{1}^{-}\left(-y_{1}^{-}y_{2}^{+}+\left(y_{2}^{+}\right)^{2}\right)\left(u_{1}-u_{2}-\frac{2i}{g}\right)}\frac{\tilde{\eta}_{1}^{2}\tilde{\eta}_{2}}{\eta_{1}\eta_{2}^{2}},\\
a_{31} & = \frac{\left(y_{1}^{-}-y_{1}^{+}\right)}{\left(y_{1}^{-}-y_{2}^{+}\right)}\frac{\tilde{\eta}_{2}}{\eta_{1}},\\
a_{32} & = \frac{\left(y_{2}^{-}-y_{2}^{+}\right)}{\left(y_{1}^{-}-y_{2}^{+}\right)}\frac{\tilde{\eta}_{1}}{\eta_{2}},\\
\end{aligned}
$

$
\begin{aligned}
a_{33} & = \frac{\left(y_{1}^{-}-y_{1}^{+}\right)\left(y_{1}^{+}-y_{2}^{+}\right)\left(-1+y_{1}^{-}y_{2}^{+}\right)}{2y_{2}^{+}\left(\left(y_{1}^{-}\right)^{2}-y_{1}^{-}y_{2}^{+}\right)\left(u_{1}-u_{2}-\frac{2i}{g}\right)}\frac{\tilde{\eta}_{2}^{2}}{\eta_{1}\eta_{2}},\\
a_{34} & = -\frac{\left(y_{2}^{-}-y_{2}^{+}\right)\left(-y_{1}^{+}+y_{2}^{+}\right)\left(-1+y_{1}^{-}y_{2}^{+}\right)}{2y_{1}^{-}\left(y_{1}^{-}-y_{2}^{+}\right)y_{2}^{+}\left(u_{1}-u_{2}-\frac{2i}{g}\right)}\frac{\tilde{\eta}_{1}\tilde{\eta}_{2}}{\eta_{2}^{2}},\\
a_{35} & = \frac{\left(y_{1}^{-}-y_{2}^{-}\right)\left(y_{2}^{-}-y_{2}^{+}\right)\left(-1+y_{1}^{-}y_{2}^{+}\right)}{2y_{2}^{+}\left(\left(y_{1}^{-}\right)^{2}-y_{1}^{-}y_{2}^{+}\right)\left(u_{1}-u_{2}-\frac{2i}{g}\right)}\frac{\tilde{\eta}_{1}^{2}}{\eta_{1}\eta_{2}},\\
a_{36} & = \frac{\left(y_{1}^{-}-y_{2}^{-}\right)\left(y_{1}^{-}-y_{1}^{+}\right)\left(-1+y_{1}^{-}y_{2}^{+}\right)}{2y_{2}^{+}\left(\left(y_{1}^{-}\right)^{2}-y_{1}^{-}y_{2}^{+}\right)\left(u_{1}-u_{2}-\frac{2i}{g}\right)}\frac{\tilde{\eta}_{1}\tilde{\eta}_{2}}{\eta_{1}^{2}},\\
a_{37} & = -\frac{i\zeta\left(y_{1}^{-}-y_{1}^{+}\right)\left(y_{2}^{-}-y_{2}^{+}\right)^{2}\left(y_{1}^{+}-y_{2}^{+}\right)}{2y_{1}^{-}y_{2}^{-}\left(y_{1}^{-}-y_{2}^{+}\right)\left(u_{1}-u_{2}-\frac{2i}{g}\right)\eta_{2}^{2}},\\
a_{38} & = \frac{i\left(y_{1}^{-}-y_{1}^{+}\right)\left(y_{1}^{+}-y_{2}^{+}\right)\tilde{\eta}_{2}^{2}}{2\zeta y_{1}^{+}\left(y_{1}^{-}-y_{2}^{+}\right)y_{2}^{+}\left(u_{1}-u_{2}-\frac{2i}{g}\right)},\\
a_{39} & = -\frac{i\zeta\left(y_{1}^{-}-y_{2}^{-}\right)\left(y_{1}^{-}-y_{1}^{+}\right)^{2}\left(y_{2}^{-}-y_{2}^{+}\right)}{2y_{1}^{-}y_{2}^{-}\left(y_{1}^{-}-y_{2}^{+}\right)\left(u_{1}-u_{2}-\frac{2i}{g}\right)\eta_{1}^{2}},\\
a_{40} & = \frac{i\left(y_{1}^{-}-y_{2}^{-}\right)\left(y_{2}^{-}-y_{2}^{+}\right)\tilde{\eta}_{1}^{2}}{2\zeta y_{1}^{+}\left(y_{1}^{-}-y_{2}^{+}\right)y_{2}^{+}\left(u_{1}-u_{2}-\frac{2i}{g}\right)},\\
a_{41} & = \frac{i\zeta\left(y_{1}^{-}-y_{1}^{+}\right)\left(y_{2}^{-}-y_{1}^{+}\right)\left(y_{2}^{-}-y_{2}^{+}\right)\left(y_{1}^{+}-y_{2}^{+}\right)}{2y_{1}^{-}y_{2}^{-}\left(y_{1}^{-}-y_{2}^{+}\right)\left(u_{1}-u_{2}-\frac{2i}{g}\right)}\frac{\tilde{\eta}_{2}}{\eta_{1}\eta_{2}^{2}},\\
a_{42} & = -\frac{i\left(y_{2}^{-}-y_{1}^{+}\right)\left(y_{1}^{+}-y_{2}^{+}\right)}{2\zeta y_{1}^{+}\left(y_{1}^{-}-y_{2}^{+}\right)y_{2}^{+}\left(u_{1}-u_{2}-\frac{2i}{g}\right)}\frac{\tilde{\eta}_{1}\tilde{\eta}_{2}^{2}}{\eta_{2}},\\
a_{43} & = \frac{i\zeta\left(y_{1}^{-}-y_{2}^{-}\right)\left(y_{1}^{-}-y_{1}^{+}\right)\left(y_{2}^{-}-y_{1}^{+}\right)\left(y_{2}^{-}-y_{2}^{+}\right)}{2y_{1}^{-}y_{2}^{-}\left(y_{1}^{-}-y_{2}^{+}\right)\left(u_{1}-u_{2}-\frac{2i}{g}\right)}\frac{\tilde{\eta}_{1}}{\eta_{1}^{2}\eta_{2}},\\
a_{44} & = -\frac{i\left(y_{1}^{-}-y_{2}^{-}\right)\left(y_{2}^{-}-y_{1}^{+}\right)}{2\zeta y_{1}^{+}\left(y_{1}^{-}-y_{2}^{+}\right)y_{2}^{+}\left(u_{1}-u_{2}-\frac{2i}{g}\right)}\frac{\tilde{\eta}_{1}^{2}\tilde{\eta}_{2}}{\eta_{1}},\\
a_{45} & = a_{46}=a_{47}=a_{48}=0.
\end{aligned}
$

\subsection{Coefficients of $K$-matrices}

We use the convenient notation for the spectral parameters in which $x^{\pm}$
and $y^{\pm}$ are the spectral parameters of the fundamental state
and two-particle bound state (respectively) in the bulk with momentum
$p$, while $x_{B}$ and $y_{B}$ are the spectral parameters of the
fundamental state and two-particle bound state respectively on the
boundary. The bulk parameters change as 
$x^{\pm} \rightarrow -x^{\mp}$, $y^{\pm} \rightarrow -y^{\mp}$ and 
$\eta \rightarrow \tilde{\eta}$ under the reflection.
The boundary parameter $\eta_{B}$ is related to the boundary spectral parameters 
as $\left|\eta_{B}\right|^{2}=-ix_{B}$
and $\left|\eta_{B}\right|^{2}=-iy_{B}$ in the cases of fundamental and
two-particle bound states respectively and changes to $\tilde{\eta}_{B}$ under the reflection.

\subsubsection*{The $K$-matrix $K^{Aa}$}

The $K$-matrix $K^{Aa}$ has the following components $k_{i}$:

$
\begin{aligned}
k_{1} & = 1,\\
k_{2} & = 1+2\frac{\left(x_{B}+x^{-}\right)\left(\left(x^{-}\right)^{2}-\left(x^{+}\right)^{2}\right)}{\left(x_{B}-x^{-}\right)x^{-}x^{+}},\\
k_{3} & = -\frac{x^{+}\left(x_{B}+x^{+}\right)}{\left(x_{B}-x^{-}\right)x^{-}}\frac{\tilde{\eta}\,\tilde{\eta}_{B}}{\eta\,\eta_{B}},\\
k_{4} & = \left(\frac{\left(x^{-}-2x^{+}\right)\left(x_{B}+x^{-}-x^{+}\right)\left(x^{-}+x^{+}\right)}{\left(x_{B}-x^{-}\right)\left(x^{-}\right)^{2}}+1\right)\frac{\tilde{\eta}\,\tilde{\eta}_{B}}{\eta\,\eta_{B}},\\
k_{5} & = \frac{\left(x_{B}x^{-}-\left(x^{+}\right)^{2}\right)}{\left(x_{B}-x^{-}\right)x^{-}}\frac{\tilde{\eta}_{B}}{\eta_{B}},\\
k_{6} & = -\frac{\left(\left(x^{-}\right)^{2}+x_{B}x^{+}\right)}{\left(x_{B}-x^{-}\right)x^{-}}\frac{\tilde{\eta}}{\eta},\\
k_{7} & = -\frac{i\zeta x_{B}\left(x_{B}+x^{-}-x^{+}\right)\left(\left(x^{-}\right)^{2}-\left(x^{+}\right)^{2}\right)}{\left(x_{B}-x^{-}\right)x^{-}\,\eta\,\eta_{B}},\\
k_{8} & = \frac{i\left(x_{B}+x^{-}-x^{+}\right)\left(x^{-}+x^{+}\right)\tilde{\eta}\,\eta_{B}}{\zeta\left(x_{B}-x^{-}\right)x^{-}},\\
k_{9} & = \frac{\left(-\left(x^{-}\right)^{2}+\left(x^{+}\right)^{2}\right)}{\left(x_{B}-x^{-}\right)x^{-}}\frac{\tilde{\eta}_{B}}{\eta},\\
k_{10} & = \frac{x_{B}\left(x^{-}+x^{+}\right)}{\left(x_{B}-x^{-}\right)x^{-}}\frac{\tilde{\eta}}{\eta_{B}}.
\end{aligned}
$

\noindent Our coefficients are in agreement with the ones found in \cite{Young1}
up to an overall factor
\begin{equation*}
N_{0}^{Aa}=\frac{x^{-}\left(x^{-}-x_{B}\right)}{x^{+}\left(x^{+}+x_{B}\right)},
\end{equation*}
which corresponds to normalization with $k_{3}=1$ that once again may be called as a physical
normalization, because in the same way as for $S$-matrix, the $K$-matrix element $k_{3}$ shouldn't have a
pole as it produces a symmetric in fermionic indices state which can't
create a bound state. Here the pole $x^{+}+x_{B}$ is responsible for the creation of bound states.
Once again, the pole is hidden in overall factors of higher order $K$-matrices because of normalization $k_{1} = 1$ that we use in calculations.

\pagebreak

\subsubsection*{\noindent The $K$-matrix $K^{Ba}$}

The $K$-matrix $K^{Ba}$ has the following components $k_{i}$:

$
\begin{aligned}
k_{1} & = 1,\\
k_{2} & = 1+3\frac{x_{B}\left(\left(y^{-}\right)^{2}-\left(y^{+}\right)^{2}\right)\left(1+\left(y^{+}\right)^{2}\right)}{2\left(x_{B}-y^{-}\right)\left(1+x_{B}y^{-}\right)\left(y^{+}\right)^{2}},\\
k_{3} & = -\frac{\left(\left(y^{-}\right)^{2}+x_{B}y^{+}\right)}{\left(x_{B}-y^{-}\right)y^{-}}\frac{\tilde{\eta}}{\eta},\\
k_{4} & = -\frac{\left(x_{B}+y^{+}\right)\left(y^{-}+x_{B}\left(y^{+}\right)^{2}\right)}{\left(x_{B}-y^{-}\right)\left(1+x_{B}y^{-}\right)y^{+}}\frac{\tilde{\eta}}{\eta},\\
k_{5} & = \frac{\left(x_{B}y^{-}-\left(y^{+}\right)^{2}\right)}{\left(x_{B}-y^{-}\right)y^{-}}\frac{\tilde{\eta}_{B}}{\eta_{B}},\\
k_{6} & = \frac{-x_{B}\left(y^{-}\right)^{4}+x_{B}y^{-}y^{+}+4\left(y^{-}\right)^{2}y^{+}+x_{B}\left(y^{-}\right)^{3}y^{+}+x_{B}\left(y^{+}\right)^{2}-2x_{B}\left(y^{-}\right)^{2}\left(y^{+}\right)^{2}}{2\left(x_{B}-y^{-}\right)\left(y^{-}\right)^{2}\left(1+x_{B}y^{-}\right)}\frac{\tilde{\eta}^2}{\eta^2},\\
k_{7} & = -\frac{y^{+}\left(x_{B}+y^{+}\right)}{\left(x_{B}-y^{-}\right)y^{-}}\frac{\tilde{\eta}\,\tilde{\eta}_{B}}{\eta\,\eta_{B}},\\
k_{8} & = \frac{\left(2x_{B}^{2}\left(y^{-}\right)^{3}-x_{B}y^{-}y^{+}+x_{B}^{2}\left(y^{-}\right)^{2}y^{+}+y^{-}\left(y^{+}\right)^{2}-x_{B}\left(y^{-}\right)^{2}\left(y^{+}\right)^{2}+2\left(y^{+}\right)^{3}\right)}{\left(x_{B}-y^{-}\right)\left(y^{-}\right)^{2}\left(1+x_{B}y^{-}\right)}\frac{\tilde{\eta}\,\tilde{\eta}_{B}}{\eta\,\eta_{B}},\\
k_{9} & = \frac{y^{+}\left(x_{B}+y^{+}\right)\left(-x_{B}\left(y^{-}\right)^{2}+y^{+}\right)}{\left(x_{B}-y^{-}\right)\left(y^{-}\right)^{2}\left(1+x_{B}y^{-}\right)}\frac{\tilde{\eta}^2\,\tilde{\eta}_{B}}{\eta^2\,\eta_{B}},\\
k_{10} & = -\frac{i\zeta x_{B}\left(\left(y^{-}\right)^{2}-\left(y^{+}\right)^{2}\right)^{2}}{2\left(x_{B}-y^{-}\right)y^{-}\left(1+x_{B}y^{-}\right)y^{+}\,\eta^{2}},\\
k_{11} & = -\frac{ix_{B}\left(y^{-}+y^{+}\right)^{2}\tilde{\eta}^{2}}{2\zeta\left(x_{B}-y^{-}\right)y^{-}\left(1+x_{B}y^{-}\right)y^{+}},\\
k_{12} & = -\frac{i\zeta x_{B}\left(x_{B}y^{-}-\left(y^{+}\right)^{2}\right)\left(\left(y^{-}\right)^{2}-\left(y^{+}\right)^{2}\right)}{\sqrt{2}\left(x_{B}-y^{-}\right)y^{-}\left(1+x_{B}y^{-}\right)y^{+}\,\eta\,\eta_{B}},\\
k_{13} & = \frac{i\left(y^{-}+y^{+}\right)\left(x_{B}y^{-}-\left(y^{+}\right)^{2}\right)\,\tilde{\eta}\,\tilde{\eta}_{B}}{\sqrt{2}\zeta\left(x_{B}-y^{-}\right)y^{-}\left(1+x_{B}y^{-}\right)y^{+}},\\
k_{14} & = \frac{x_{B}\left(x_{B}\left(y^{-}\right)^{2}-y^{+}\right)\left(y^{-}+y^{+}\right)}{\sqrt{2}\left(y^{-}\right)^{2}\left(-x_{B}+y^{-}\right)\left(1+x_{B}y^{-}\right)}\frac{\tilde{\eta}^2}{\eta\,\eta_{B}},\\
k_{15} & = \frac{\left(x_{B}\left(y^{-}\right)^{2}-y^{+}\right)\left(\left(y^{-}\right)^{2}-\left(y^{+}\right)^{2}\right)}{\sqrt{2}\left(x_{B}-y^{-}\right)\left(y^{-}\right)^{2}\left(1+x_{B}y^{-}\right)}\frac{\tilde{\eta}\,\eta_{B}}{\eta^2},\\
k_{16} & = -\frac{i\left(x_{B}+y^{+}\right)\left(y^{-}+y^{+}\right)}{\sqrt{2}\zeta y^{-}\left(-x_{B}+y^{-}\right)\left(1+x_{B}y^{-}\right)}\frac{\tilde{\eta}^2\,\tilde{\eta}_{B}}{\eta},\\
\end{aligned}
$

$
\begin{aligned}
k_{17} & = \frac{i\zeta x_{B}\left(x_{B}+y^{+}\right)\left(-\left(y^{-}\right)^{2}+\left(y^{+}\right)^{2}\right)}{\sqrt{2}\left(x_{B}-y^{-}\right)y^{-}\left(1+x_{B}y^{-}\right)}\frac{\tilde{\eta}}{\eta^2\,\eta_{B}},\\
k_{18} & = \frac{x_{B}\left(y^{-}+y^{+}\right)}{\sqrt{2}\left(x_{B}-y^{-}\right)y^{-}}\frac{\tilde{\eta}}{\eta_{B}},\\
k_{19} & = \frac{\left(\left(y^{-}\right)^{2}-\left(y^{+}\right)^{2}\right)}{\sqrt{2}y^{-}\left(-x_{B}+y^{-}\right)}\frac{\tilde{\eta}_{B}}{\eta}.
\end{aligned}
$

\subsubsection*{The K-Matrix $K^{Ab}$}

The $K$-matrix $K^{Ab}$ has the following components $k_{i}$:

$
\begin{aligned}
k_{1} & = 1,\\
k_{2} & = 1-\frac{3}{2}\frac{\left(x^{-}+x^{+}\right)\left(\left(x^{-}\right)^{2}+y_{B}^{2}\left(x^{+}\right)^{2}\right)}{\left(y_{B}-x^{-}\right)x^{-}x^{+}\left(-1+y_{B}x^{+}\right)},\\
k_{3} & = \frac{\left(y_{B}x^{-}-\left(x^{+}\right)^{2}\right)}{\left(y_{B}-x^{-}\right)x^{-}}\frac{\tilde{\eta}_{B}}{\eta_{B}},\\
k_{4} & = -\frac{\left(y_{B}+x^{+}\right)\left(x^{-}+y_{B}\left(x^{+}\right)^{2}\right)}{\left(y_{B}-x^{-}\right)x^{-}\left(-1+y_{B}x^{+}\right)}\frac{\tilde{\eta}_{B}}{\eta_{B}},\\
k_{5} & = -\frac{\left(\left(x^{-}\right)^{2}+y_{B}x^{+}\right)}{\left(y_{B}-x^{-}\right)x^{-}}\frac{\tilde{\eta}}{\eta},\\
k_{6} & = \frac{x^{+}\left(y_{B}^{2}\left(x^{-}\right)^{3}-2y_{B}x^{-}x^{+}-y_{B}^{2}\left(x^{-}\right)^{2}x^{+}-x^{-}\left(x^{+}\right)^{2}-2y_{B}\left(x^{-}\right)^{2}\left(x^{+}\right)^{2}+\left(x^{+}\right)^{3}\right)\tilde{\eta}_{B}^{2}}{2\left(y_{B}-x^{-}\right)\left(x^{-}\right)^{3}\left(-1+y_{B}x^{+}\right)\eta_{B}^{2}},\\
k_{7} & = -\frac{x^{+}\left(y_{B}+x^{+}\right)}{\left(y_{B}-x^{-}\right)x^{-}}\frac{\tilde{\eta}\,\tilde{\eta}_{B}}{\eta\,\eta_{B}},\\
k_{8} & = \frac{x^{+}\left(\left(x^{-}\right)^{2}x^{+}-y_{B}\left(x^{-}\right)^{2}-2y_{B}\left(x^{-}\right)^{4}-y_{B}^{2}\left(x^{-}\right)^{2}x^{+}+2y_{B}\left(x^{+}\right)^{2}+y_{B}\left(x^{-}\right)^{2}\left(x^{+}\right)^{2}\right)\tilde{\eta}\,\tilde{\eta}_{B}}{\left(y_{B}-x^{-}\right)\left(x^{-}\right)^{3}\left(-1+y_{B}x^{+}\right)\eta\,\eta_{B}},\\
k_{9} & = \frac{\left(x^{+}\right)^{2}\left(y_{B}+x^{+}\right)\left(-y_{B}\left(x^{-}\right)^{2}+x^{+}\right)}{\left(y_{B}-x^{-}\right)\left(x^{-}\right)^{3}\left(-1+y_{B}x^{+}\right)}\frac{\tilde{\eta}\,\tilde{\eta}_{B}^2}{\eta\,\eta_{B}^2},\\
k_{10} & = -\frac{i\zeta y_{B}^{2}\left(x^{-}-x^{+}\right)\left(x^{-}+x^{+}\right)^{2}}{2\left(y_{B}-x^{-}\right)\left(x^{-}\right)^{2}\left(-1+y_{B}x^{+}\right)\eta_{B}^{2}},\\
k_{11} & = -\frac{i\left(x^{-}-x^{+}\right)\left(x^{-}+x^{+}\right)^{2}\tilde{\eta}_{B}^{2}}{2\zeta\left(y_{B}-x^{-}\right)\left(x^{-}\right)^{2}\left(-1+y_{B}x^{+}\right)},\\
k_{12} & = \frac{i\zeta y_{B}\left(\left(x^{-}\right)^{2}+y_{B}x^{+}\right)\left(\left(x^{-}\right)^{2}-\left(x^{+}\right)^{2}\right)}{\sqrt{2}\left(y_{B}-x^{-}\right)\left(x^{-}\right)^{2}\left(-1+y_{B}x^{+}\right)\eta\,\eta_{B}},\\
\end{aligned}
$

$
\begin{aligned}
k_{13} & = \frac{i\left(x^{-}+x^{+}\right)\left(\left(x^{-}\right)^{2}+y_{B}x^{+}\right)\tilde{\eta}\,\tilde{\eta}_{B}}{\sqrt{2}\zeta\left(x^{-}\right)^{2}\left(-y_{B}+x^{-}\right)\left(-1+y_{B}x^{+}\right)},\protect\\
k_{14} & = \frac{\left(y_{B}\left(x^{-}\right)^{2}-x^{+}\right)x^{+}\left(\left(x^{-}\right)^{2}-\left(x^{+}\right)^{2}\right)}{\sqrt{2}\left(x^{-}\right)^{3}\left(-y_{B}+x^{-}\right)\left(-1+y_{B}x^{+}\right)}\frac{\tilde{\eta}_{B}^2}{\eta\,\eta_{B}},\\
k_{15} & = \frac{y_{B}\left(y_{B}\left(x^{-}\right)^{2}-x^{+}\right)x^{+}\left(x^{-}+x^{+}\right)}{\sqrt{2}\left(y_{B}-x^{-}\right)\left(x^{-}\right)^{3}\left(-1+y_{B}x^{+}\right)}\frac{\tilde{\eta}\,\tilde{\eta}_{B}}{\eta^2_{B}},\\
k_{16} & = \frac{ix^{+}\left(y_{B}+x^{+}\right)\left(x^{-}+x^{+}\right)}{\sqrt{2}\zeta\left(x^{-}\right)^{2}\left(-y_{B}+x^{-}\right)\left(-1+y_{B}x^{+}\right)}\frac{\tilde{\eta}\,\tilde{\eta}_{B}^2}{\eta_{B}},\\
k_{17} & = \frac{i\zeta y_{B}x^{+}\left(y_{B}+x^{+}\right)\left(\left(x^{-}\right)^{2}-\left(x^{+}\right)^{2}\right)}{\sqrt{2}\left(y_{B}-x^{-}\right)\left(x^{-}\right)^{2}\left(-1+y_{B}x^{+}\right)}\frac{\tilde{\eta}_{B}}{\eta\,\eta_{B}^2},\\
k_{18} & = \frac{\left(\left(x^{-}\right)^{2}-\left(x^{+}\right)^{2}\right)}{\sqrt{2}x^{-}\left(-y_{B}+x^{-}\right)}\frac{\tilde{\eta}_{B}}{\eta},\\
k_{19} & = \frac{y_{B}\left(x^{-}+x^{+}\right)}{\sqrt{2}\left(y_{B}-x^{-}\right)x^{-}}\frac{\tilde{\eta}}{\eta_{B}}.
\end{aligned}
$

\subsubsection*{The $K$-matrix $K^{Bb}$}

The $K$-matrix $K^{Bb}$ has the following components $k_{i}$:

$
\begin{aligned}
k_{1} & = 1,\\
k_{2} & = -\frac{\left(y_{B}+y^{+}\right)\left(y^{-}+y_{B}\left(y^{+}\right)^{2}\right)}{\left(y_{B}-y^{-}\right)\left(1+y_{B}y^{-}\right)\left(y^{+}\right)^{3}\,O_{1}\,}\\
 & \quad\times\left(3y_{B}\left(y^{-}\right)^{2}+\left(-y_{B}+y^{-}\left(-2+2y_{B}^{2}+y_{B}y^{-}\right)\right)\left(y^{+}\right)^{2}-3y_{B}\left(y^{+}\right)^{4}\right),\\
k_{3} & = -\frac{1}{\left(y_{B}-y^{-}\right)\left(1+y_{B}y^{-}\right)\left(y^{+}\right)^{3}\,O_{1}\,}\\
 & \quad\times\Bigl(-y_{B}\left(y^{+}\right)^{4}\left(1+y_{B}^{2}+2\left(y^{+}\right)^{2}-2y_{B}\left(y^{+}\right)^{3}\right)\\
 & \quad\qquad+y_{B}\left(y^{-}\right)^{3}\left(-2y_{B}+2y_{B}^{2}y^{+}+\left(1+y_{B}^{2}\right)\left(y^{+}\right)^{3}\right)\\
 & \quad\qquad+\left(y^{-}\right)^{2}\left(y^{+}\right)^{2}\left(-2y_{B}\left(-2+y_{B}^{2}\right)+y^{+}\left(1-4y_{B}^{2}+y_{B}^{4}+\left(y_{B}+y_{B}^{3}\right)y^{+}\right)\right)\\
 & \quad\qquad-y^{-}\left(y^{+}\right)^{3}\left(y_{B}+y_{B}^{3}+y^{+}\left(1-4y_{B}^{2}+y_{B}^{4}+\left(-2y_{B}+4y_{B}^{3}\right)y^{+}\right)\right)\Bigr),\\
k_{4} & = -\frac{y^{+}\left(y_{B}+y^{+}\right)}{\left(y_{B}-y^{-}\right)y^{-}}\frac{\tilde{\eta}\,\tilde{\eta}_{B}}{\eta\,\eta_{B}},\\
\end{aligned}
$

$
\begin{aligned}
k_{5} & = -\frac{\tilde{\eta}\,\tilde{\eta}_{B}}{\left(y_{B}-y^{-}\right)\left(y^{-}\right)^{2}\left(1+y_{B}y^{-}\right)y^{+}O_{1}\,\eta\,\eta_{B}}\\
 & \quad\times\left(4y_{B}^{3}\left(y^{-}\right)^{4}\left(-1+y_{B}y^{+}\right)+y^{-}\left(y^{+}\right)^{3}\left(y_{B}\left(-5+y_{B}^{2}\right)+y^{+}\left(3+7y_{B}^{2}-2y_{B}y^{+}\right)\right)\right.\\
 & \quad\qquad-4\left(y^{+}\right)^{5}\left(-1+y_{B}y^{+}\right)+y_{B}\left(y^{-}\right)^{3}y^{+}\left(-2y_{B}^{2}+y^{+}\left(7y_{B}+3y_{B}^{3}+y^{+}-5y_{B}^{2}y^{+}\right)\right)\\
 & \quad\qquad+\left.\left(y^{-}\right)^{2}y^{+}\left(2y_{B}^{2}+y^{+}\left(y^{+}\left(1+6y_{B}^{2}+y_{B}^{4}-y_{B}y^{+}\left(1+y_{B}^{2}-2y_{B}y^{+}\right)\right)-y_{B}\left(1+y_{B}^{2}\right)\right)\right)\right),\\
k_{6} & = -\frac{\left(y_{B}+y^{+}\right)O_{2}}{\left(y_{B}-y^{-}\right)y^{-}O_{1}}\frac{\tilde{\eta}\,\tilde{\eta}_{B}}{\eta\,\eta_{B}},\\
k_{7} & = \frac{\left(y_{B}+y^{+}\right)\,\tilde{\eta}\,\tilde{\eta}_{B}}{\left(y_{B}-y^{-}\right)\left(y^{-}\right)^{2}\left(1+y_{B}y^{-}\right)y^{+}O_{1}\,\eta\,\eta_{B}},\\
 & \quad\times\left(4y_{B}^{2}\left(y^{-}\right)^{4}+y^{-}\left(y^{+}\right)^{3}\left(3-3y_{B}^{2}+2y_{B}y^{+}\right)+y_{B}\left(y^{-}\right)^{3}y^{+}\left(2y_{B}+3\left(-1+y_{B}^{2}\right)y^{+}\right)\right.\\
 & \quad\qquad\left.+4y_{B}\left(y^{+}\right)^{5}+\left(y^{-}\right)^{2}y^{+}\left(1+y_{B}y^{+}\right)\left(-2y_{B}+y^{+}\left(1+y_{B}^{2}-2y_{B}y^{+}\right)\right)\right),\\
k_{8} & = -\frac{\left(y_{B}\left(y^{-}\right)^{2}-y^{+}\right)\left(y^{+}\right)^{2}\left(y_{B}+y^{+}\right)}{\left(y_{B}-y^{-}\right)\left(y^{-}\right)^{5}\left(-1+y_{B}y^{+}\right)O_{1}}\frac{\tilde{\eta}^2\,\tilde{\eta}_{B}^2}{\eta^2\,\eta_{B}^2}\\
 & \quad\times\left(y_{B}\left(y^{-}\right)^{4}-y_{B}\left(y^{+}\right)^{2}+\left(y^{-}\right)^{2}\left(-y_{B}+y^{+}\left(-2+2y_{B}^{2}+y_{B}y^{+}\right)\right)\right),\\
k_{9} & = \frac{y_{B}y^{-}-\left(y^{+}\right)^{2}}{\left(y_{B}-y^{-}\right)y^{-}}\frac{\tilde{\eta}_{B}}{\eta_{B}},\\
k_{10} & = -\frac{\left(y^{-}\right)^{2}+y_{B}y^{+}}{y^{-}\left(y_{B}-y^{-}\right)}\frac{\tilde{\eta}}{\eta},\\
k_{11} & = \frac{2\left(y_{B}+y^{+}\right)\left(y_{B}y^{-}-\left(y^{+}\right)^{2}\right)\left(y^{-}+y_{B}\left(y^{+}\right)^{2}\right)\tilde{\eta}_{B}}{\left(y_{B}-y^{-}\right)y^{-}y^{+}O_{1}\,\eta_{B}},\\
k_{12} & = \frac{2\left(\left(y^{-}\right)^{2}+y_{B}y^{+}\right)}{\left(y_{B}-y^{-}\right)\left(y^{-}\right)^2y^{+}O_{1}}\frac{\tilde{\eta}}{\eta}\\
 & \quad\times\left(-y_{B}\left(y^{-}\right)^{2}-y^{-}\left(y_{B}+y^{-}\right)y^{+}+\left(y_{B}+y^{-}+y_{B}\left(y^{-}\right)^{2}\right)\left(y^{+}\right)^{2}-2y_{B}y^{-}\left(y^{+}\right)^{3}\right),\\
k_{13} & = \frac{\left(y_{B}y^{-}-\left(y^{+}\right)^{2}\right)\tilde{\eta}_{B}^{2}}{\left(y_{B}-y^{-}\right)\left(y^{-}\right)^{3}\left(1+y_{B}y^{-}\right)y^{+}O_{1}\,\eta_{B}^{2}},\\
 & \quad\times\Bigl(-y_{B}^{3}\left(y^{-}\right)^{3}+y_{B}^{4}\left(y^{-}\right)^{3}y^{+}+y_{B}y^{-}\left(2+y_{B}^{2}+y_{B}y^{-}\left(5+y_{B}^{2}+y_{B}y^{-}\right)\right)\left(y^{+}\right)^{2}\\
 & \quad\qquad+y^{-}\left(1+y_{B}y^{-}\left(2-2y_{B}^{2}+y_{B}y^{-}\right)\right)\left(y^{+}\right)^{3}\\
 & \quad\qquad-\left(1+y_{B}^{2}+y_{B}\left(2+y_{B}^{2}\right)y^{-}\right)\left(y^{+}\right)^{4}+y_{B}^{2}y^{-}\left(y^{+}\right)^{5}\Bigr),\\
k_{14} & = -\frac{\left(\left(y^{-}\right)^{2}+y_{B}y^{+}\right)\left(y_{B}\left(y^{-}\right)^{4}-y_{B}\left(y^{+}\right)^{2}+\left(y^{-}\right)^{2}\left(-y_{B}+y^{+}\left(-2+2y_{B}^{2}+y_{B}y^{+}\right)\right)\right)\,\tilde{\eta}^2}{\left(y_{B}-y^{-}\right)\left(y^{-}\right)^{3}O_{1}\,\eta^2},\\
k_{15} & = -\frac{2\left(y_{B}\left(y^{-}\right)^{2}-y^{+}\right)y^{+}\left(y_{B}+y^{+}\right)\left(\left(y^{-}\right)^{2}+y_{B}y^{+}\right)\,\tilde{\eta}^2\,\tilde{\eta}_{B}}{\left(y_{B}-y^{-}\right)\left(y^{-}\right)^{3}O_{1}\eta^2\,\eta_{B}},\\
\end{aligned}
$

$
\begin{aligned}
k_{16} & = -\frac{2\left(y_{B}\left(y^{-}\right)^{2}-y^{+}\right)\left(y^{+}\right)^{2}\left(y_{B}+y^{+}\right)}{\left(y_{B}-y^{-}\right)\left(y^{-}\right)^{3}\left(-1+y_{B}y^{+}\right)O_{1}}\frac{\tilde{\eta}\,\tilde{\eta}_{B}^{2}}{\eta\,\eta_{B}^{2}}\\
 & \quad\times\left(y_{B}+y^{-}+y_{B}\left(y^{-}\right)^{2}-\left(1+y_{B}y^{-}\right)y^{+}+y_{B}\left(y^{+}\right)^{2}\right),\\
k_{17} & = -\frac{\zeta^{2}y_{B}^{2}\left(y_{B}+y^{+}\right)\left(\left(y^{-}\right)^{2}+y_{B}y^{+}\right)\left(\left(y^{-}\right)^{2}-\left(y^{+}\right)^{2}\right)^{2}}{\left(y_{B}-y^{-}\right)\left(y^{-}\right)^{3}\left(-1+y_{B}y^{+}\right)O_{1}\,\eta^{2}\,\eta_{B}^{2}},\\
k_{18} & = \frac{\left(y_{B}+y^{+}\right)\left(y^{-}+y^{+}\right)^{2}\left(y_{B}y^{-}-\left(y^{+}\right)^{2}\right)\tilde{\eta}^{2}\,\tilde{\eta}_{B}^{2}}{\zeta^{2}\left(y_{B}-y^{-}\right)\left(y^{-}\right)^{2}\left(1+y_{B}y^{-}\right)y^{+}O_{1}\,},\\
k_{19} & = -\frac{i\zeta y_{B}\left(y_{B}+y^{+}\right)\left(y_{B}y^{-}-\left(y^{+}\right)^{2}\right)\left(\left(y^{-}\right)^{2}-\left(y^{+}\right)^{2}\right)\left(y^{-}+y_{B}\left(y^{+}\right)^{2}\right)}{\left(y_{B}-y^{-}\right)y^{-}\left(1+y_{B}y^{-}\right)\left(y^{+}\right)^2 O_{1}\,\eta\,\eta_{B}},\\
k_{20} & = \frac{i\left(y_{B}+y^{+}\right)\left(y^{-}+y^{+}\right)\left(y_{B}y^{-}-\left(y^{+}\right)^{2}\right)\left(y^{-}+y_{B}\left(y^{+}\right)^{2}\right)\tilde{\eta}\,\tilde{\eta}_{B}}{\zeta\left(y_{B}-y^{-}\right)y^{-}\left(1+y_{B}y^{-}\right)\left(y^{+}\right)^2 O_{1}\,},\\
k_{21} & = -\frac{i\left(y_{B}\left(y^{-}\right)^{2}-y^{+}\right)\left(y_{B}+y^{+}\right)\left(y^{-}+y^{+}\right)\left(y_{B}y^{-}-\left(y^{+}\right)^{2}\right)\tilde{\eta}^2\,\tilde{\eta}_{B}^{2}}{\zeta\left(y_{B}-y^{-}\right)\left(y^{-}\right)^{3}\left(1+y_{B}y^{-}\right)\,O_{1}\,\eta\,\eta_{B}},\\
k_{22} & = -\frac{i\zeta y_{B}\left(y_{B}\left(y^{-}\right)^{2}-y^{+}\right)y^{+}\left(y_{B}+y^{+}\right)\left(\left(y^{-}\right)^{2}+y_{B}y^{+}\right)\left(\left(y^{-}\right)^{2}-\left(y^{+}\right)^{2}\right)\tilde{\eta}\,\tilde{\eta}_{B}}{\left(y_{B}-y^{-}\right)\left(y^{-}\right)^{4}\left(-1+y_{B}y^{+}\right)O_{1}\,\eta^2\,\eta_{B}^{2}},\\
k_{23} & = \frac{i\zeta y_{B}\left(-1+y_{B}y^{+}\right)\left(-y_{B}y^{-}+\left(y^{+}\right)^{2}\right)^{2}\left(\left(y^{-}\right)^{2}-\left(y^{+}\right)^{2}\right)}{\left(y_{B}-y^{-}\right)y^{-}\left(1+y_{B}y^{-}\right)\left(y^{+}\right)^{2}O_{1}\,\eta\,\eta_{B}},\\
k_{24} & = \frac{i\left(y^{-}+y^{+}\right)\left(-1+y_{B}y^{+}\right)\left(y_{B}y^{-}-\left(y^{+}\right)^{2}\right)^{2}\tilde{\eta}\,\tilde{\eta}_{B}}{\zeta\left(y_{B}-y^{-}\right)y^{-}\left(1+y_{B}y^{-}\right)\left(y^{+}\right)^{2}O_{1}\,},\\
k_{25} & = -\frac{\left(y_{B}\left(y^{-}\right)^{2}-y^{+}\right)\left(\left(y^{-}\right)^{2}-\left(y^{+}\right)^{2}\right)^{2}\tilde{\eta}_{B}^{2}}{\left(y_{B}-y^{-}\right)\left(y^{-}\right)^{3}O_{1}\,\eta^{2}},\\
k_{26} & = \frac{y_{B}^{2}\left(y^{-}+y^{+}\right)^{2}\tilde{\eta}^{2}}{\left(y_{B}-y^{-}\right)\left(y^{-}\right)^{3}\left(1+y_{B}y^{-}\right)y^{+}O_{1}\,\eta_{B}^{2}}\\
 & \quad\times\left(y_{B}\left(y^{-}\right)^{4}y^{+}+\left(y^{+}\right)^{2}+y_{B}y^{-}\left(y^{+}\right)^{2}-\left(y^{-}\right)^{3}\left(y_{B}+y^{+}\left(-1+y_{B}y^{+}\right)\right)\right),\\
k_{27} & = -\frac{\left(\left(y^{-}\right)^{2}\left(2+y_{B}^{2}+y_{B}y^{-}\right)+\left(y_{B}+y^{-}+2y_{B}^{2}y^{-}\right)y^{+}\right)\left(\left(y^{-}\right)^{2}-\left(y^{+}\right)^{2}\right)\tilde{\eta}_{B}}{\left(y_{B}-y^{-}\right)\left(y^{-}\right)^{2}O_{1}\,\eta},\\
k_{28} & = \frac{y_{B}\left(y_{B}+y^{+}\right)\left(y^{-}+y^{+}\right)\left(y^{-}+y_{B}\left(y^{+}\right)^{2}\right)\tilde{\eta}}{\left(y_{B}-y^{-}\right)y^{-}y^{+}O_{1}\,\eta_{B}},\\
k_{29} & = \frac{y^{+}\left(y_{B}+y^{+}\right)\left(-y_{B}\left(y^{-}\right)^{2}+y^{+}\right)\left(\left(y^{-}\right)^{2}-\left(y^{+}\right)^{2}\right)\tilde{\eta}\,\tilde{\eta}_{B}^{2}}{\left(y_{B}-y^{-}\right)\left(y^{-}\right)^{3}O_{1}\,\eta^2\,\eta_{B}},\\
k_{30} & = \frac{y_{B}\left(y_{B}\left(y^{-}\right)^{2}-y^{+}\right)y^{+}\left(y_{B}+y^{+}\right)\left(y^{-}+y^{+}\right)\tilde{\eta}^2\,\tilde{\eta}_{B}}{\left(y_{B}-y^{-}\right)\left(y^{-}\right)^{3}O_{1}\,\eta\,\eta_{B}^{2}},\\
\end{aligned}
$

$
\begin{aligned}
k_{31} & = \frac{\left(y^{+}\right)^{2}-\left(y^{-}\right)^{2}}{\left(y_{B}-y^{-}\right)y^{-}}\frac{\tilde{\eta}_{B}}{\eta},\\
k_{32} & = \frac{y_{B}\left(y^{-}+y^{+}\right)}{\left(y_{B}-y^{-}\right)y^{-}}\frac{\tilde{\eta}}{\eta_{B}},\\
k_{33} & = -\frac{\left(y^{-}+y^{+}\right)\left(y_{B}y^{-}-\left(y^{+}\right)^{2}\right)}{y_{B}\left(y_{B}-y^{-}\right)\left(y^{-}\right)^{2}\left(1+y_{B}y^{-}\right)y^{+}O_{1}}\frac{\tilde{\eta}_{B}^{2}}{\eta\,\eta_{B}}\\
 & \quad\times\left(\left(y^{+}\right)^{3}-y_{B}\left(y^{+}\right)^{4}+y_{B}y^{-}y^{+}\left(y_{B}+y^{+}\right)^{2}+y_{B}^{3}\left(y^{-}\right)^{2}\left(-1+y_{B}y^{+}\right)\right),\\
k_{34} & = \frac{y_{B}\left(y^{-}+y^{+}\right)\left(y_{B}y^{-}-\left(y^{+}\right)^{2}\right)}{\left(y_{B}-y^{-}\right)\left(y^{-}\right)^{3}\left(1+y_{B}y^{-}\right)y^{+}O_{1}}\frac{\tilde{\eta}\,\tilde{\eta}_{B}}{\eta_{B}^{2}}\\
 & \quad\times\left(y_{B}\left(y^{-}\right)^{4}y^{+}+\left(y^{+}\right)^{2}+y_{B}y^{-}\left(y^{+}\right)^{2}-\left(y^{-}\right)^{3}\left(y_{B}+y^{+}\left(-1+y_{B}y^{+}\right)\right)\right),\\
k_{35} & = -\frac{y_{B}\left(y_{B}\left(y^{-}\right)^{2}-y^{+}\right)\left(y^{-}+y^{+}\right)\left(-1+y_{B}y^{+}\right)\left(y_{B}y^{-}-\left(y^{+}\right)^{2}\right)\tilde{\eta}^2}{\left(y_{B}-y^{-}\right)\left(y^{-}\right)^{2}\left(1+y_{B}y^{-}\right)y^{+}O_{1}\,\eta\,\eta_{B}},\\
k_{36} & = -\frac{\left(y_{B}\left(y^{-}\right)^{2}-y^{+}\right)\left(\left(y^{-}\right)^{2}+y_{B}y^{+}\right)\left(\left(y^{-}\right)^{2}-\left(y^{+}\right)^{2}\right)\tilde{\eta}\,\tilde{\eta}_{B}}{\left(y_{B}-y^{-}\right)\left(y^{-}\right)^{3}O_{1}\,\eta^2},\\
k_{37} & = -\frac{i\zeta y_{B}^{2}\left(y^{-}-y^{+}\right)\left(y^{-}+y^{+}\right)^{2}}{\left(y_{B}-y^{-}\right)\left(y^{-}\right)^{2}\left(1+y_{B}y^{-}\right)\left(y^{+}\right)^{2}O_{1}\,\eta_{B}^{2}}\\
 & \quad\times\left(y_{B}\left(y^{-}\right)^{3}y^{+}+\left(y^{+}\right)^{3}+y_{B}y^{-}\left(y^{+}\right)^{3}-\left(y^{-}\right)^{2}\left(y_{B}+y^{+}\left(-1+y_{B}y^{+}\right)\right)\right),\\
k_{38} & = -\frac{i\left(y^{-}-y^{+}\right)\left(y^{-}+y^{+}\right)^{2}\left(y_{B}y^{-}-\left(y^{+}\right)^{2}\right)\left(-y_{B}+y^{+}\left(1+y_{B}y^{-}-y_{B}y^{+}\right)\right)\tilde{\eta}_{B}^{2}}{\zeta y_{B}\left(y_{B}-y^{-}\right)y^{-}\left(1+y_{B}y^{-}\right)\left(y^{+}\right)^{2}O_{1}\,},\\
k_{39} & = \frac{i\zeta y_{B}\left(\left(y^{-}\right)^{2}+y_{B}y^{+}\right)\left(\left(y^{-}\right)^{2}-\left(y^{+}\right)^{2}\right)^{2}}{\left(y_{B}-y^{-}\right)\left(y^{-}\right)^{2}y^{+}O_{1}\,\eta^{2}},\\
\end{aligned}
$

$
\begin{aligned}
k_{40} & = -\frac{iy_{B}\left(y^{-}+y^{+}\right)^{2}\left(-1+y_{B}y^{+}\right)\left(y_{B}y^{-}-\left(y^{+}\right)^{2}\right)\tilde{\eta}^{2}}{\zeta\left(y_{B}-y^{-}\right)y^{-}\left(1+y_{B}y^{-}\right)\left(y^{+}\right)^{2}O_{1}\,},\\
k_{41} & = \frac{i\zeta y_{B}\left(1+y_{B}y^{-}\right)y^{+}\left(y_{B}+y^{+}\right)\left(\left(y^{-}\right)^{2}+y_{B}y^{+}\right)\left(\left(y^{-}\right)^{2}-\left(y^{+}\right)^{2}\right)\tilde{\eta}_{B}}{\left(y_{B}-y^{-}\right)\left(y^{-}\right)^{3}\left(-1+y_{B}y^{+}\right)O_{1}\,\eta\,\eta_{B}^{2}},\\
k_{42} & = -\frac{i\left(y_{B}+y^{+}\right)\left(y^{-}+y^{+}\right)\left(-y_{B}y^{-}+\left(y^{+}\right)^{2}\right)\tilde{\eta}\,\tilde{\eta}_{B}^{2}}{\zeta\left(y_{B}-y^{-}\right)\left(y^{-}\right)^{2}\,O_{1}\,\eta_{B}},\\
k_{43} & = \frac{i\zeta y_{B}\left(y_{B}+y^{+}\right)\left(\left(y^{-}\right)^{2}+y_{B}y^{+}\right)\left(\left(y^{-}\right)^{2}-\left(y^{+}\right)^{2}\right)\tilde{\eta}}{\left(y_{B}-y^{-}\right)\left(y^{-}\right)^{2}O_{1}\,\eta^2\,\eta_{B}},\\
k_{44} & = -\frac{i\left(y_{B}+y^{+}\right)\left(y^{-}+y^{+}\right)\left(\left(y^{-}\right)^{2}+y_{B}y^{+}\right)\tilde{\eta}^2\,\tilde{\eta}_{B}}{\zeta\left(y_{B}-y^{-}\right)\left(y^{-}\right)^{2}O_{1}\,\eta},\\
k_{45} & = k_{46}=k_{47}=k_{48}=0,
\end{aligned}
$
\end{flushleft}

where we have defined
\begin{eqnarray*}
O_{1} & = & \frac{1}{2}\left(2y_{B}+y^{-}-y_{B}^{2}y^{-}+2y_{B}\left(y^{-}\right)^{2}-y^{+}\left(3+y_{B}y^{-}\right)+2y_{B}\left(y^{+}\right)^2\right),\\
O_{2} & = & \left(\left(y^{+}\right)^{2}\left(-1+y_{B}^{2}-2y_{B}y^{+}\right)+y^{-}\left(2y_{B}+y^{+}-y_{B}^{2}y^{+}\right)\right).
\end{eqnarray*}


\end{document}